\documentclass[%
 reprint,
 amsmath,amssymb,
 aps,
showkeys,floatfix, pra, aps, amsmath, amssymb, superscriptaddress, twocolumn, longbibliography]{revtex4-2}

\usepackage{graphicx}
\usepackage{dcolumn}
\usepackage{bm}
\usepackage{multirow}
\usepackage{array}
\usepackage{makecell}
\usepackage{physics}
\usepackage{hyperref}

\begin{document}

\preprint{APS/123-QED}

\title{Determination of atomic number density in MEMS vapor cells via Single-Pass Absorption Spectroscopy (SPAS)}
\author{Sumit Achar}
\thanks{These authors contributed equally to this work.}
\affiliation{Department of Physics, Indian Institute of Technology Tirupati, Yerpedu-517619, Andhra Pradesh, India.}

\author{Shivam Sinha}
\thanks{These authors contributed equally to this work.}
\affiliation{Department of Physics, Indian Institute of Technology Tirupati, Yerpedu-517619, Andhra Pradesh, India.}

\author{Ezhilarasan M}
\affiliation{Department of Physics, Indian Institute of Technology Tirupati, Yerpedu-517619, Andhra Pradesh, India.}

\author{Chandankumar R}
\affiliation{Department of Physics, Indian Institute of Technology Tirupati, Yerpedu-517619, Andhra Pradesh, India.}

\author{Arijit Sharma}
\email{arijit@iittp.ac.in}
\affiliation{Department of Physics, Indian Institute of Technology Tirupati, Yerpedu-517619, Andhra Pradesh, India.}
\affiliation{Center for Atomic, Molecular, and Optical Sciences and Technologies, Indian Institute of Technology Tirupati, Yerpedu-517619, Andhra Pradesh, India.}




\date{\today}

\begin{abstract}
Micro-electro-mechanical systems (MEMS)-based (chip-scale) alkali vapor cells are key components in emerging quantum technologies, where device performance critically depends on the atomic number density. Thus, it is important to have an accurate estimate of the atomic number density in MEMS-based alkali vapor cells to optimize light-matter interactions and design efficient quantum sensing systems. Here, a quantitatively validated method is presented for determining the rubidium (Rb) atomic number density in warm vapor using Single-Pass Absorption Spectroscopy (SPAS). The absolute transmission spectra are measured and modeled using the 780.24~nm and 420.29~nm transitions in Rb-filled MEMS vapor. The theoretical model employs a density-matrix formalism within the Lindblad framework and incorporates directly measurable experimental parameters, such as laser beam power, diameter, and cell temperature. The model explicitly accounts for optical pumping, Doppler broadening, and transit-time broadening effects and exhibits quantitative agreement ($> 99\%$) with experimental spectra over a broad range of temperatures (293-353~K), laser probe powers of approximately 10~$\mu$W-100~$\mu$W at the 780.24~nm transition and 8~$\mu$W-80~$\mu$W at the 420.29~nm transition, and cell lengths (2--100~mm). This method demonstrates a practical and reliable approach for determining the density of alkali vapor cells for quantum sensing, metrology and quantum communication applications.

\end{abstract}

\keywords{Single-Pass Absorption Spectroscopy, number density, Density matrix, MEMS}
\maketitle

\section{Introduction}\label{sec1}
With the growing demand for portable quantum devices for use in precision timing, sensitive magnetometry \cite{fu2020sensitive}, electric-field sensing \cite{sedlacek2012microwave, cox2018quantum, chen2025electric}, and entangled photon-pair generation \cite{kwiat1995new,reimer2016generation}, etc., micro-fabricated chip-scale vapor cells based on micro-electro-mechanical systems (MEMS) technology \cite{MEMS} have become key components of field-deployable quantum devices for sensing, communication, and precision timing. These miniaturized vapor cells enable wide applications in quantum technologies, including compact atomic clocks \cite{HASEGAWA2011594}, high-precision magnetometers \cite{budker2007optical}, frequency-metrology platforms \cite{udem2002optical}, and integrated photonic–atomic sensors \cite{demtroder2013laser}. Their compact form factor, low power consumption, greater reliability, and low cost for scalable fabrication make them particularly attractive for next-generation quantum applications. Most quantum devices rely on the well-defined optical and spectroscopic properties of alkali vapors, enabling applications ranging from inertial navigation \cite{aggarwal2010mems} and timing synchronization \cite{4107685} to biomedical applications through ultra-low magnetic field detection \cite{PhysRevApplied.14.011002}.\\

In chip-scale quantum devices, particularly MEMS-based vapor cells, the intrinsically short optical path length makes the optical depth extremely sensitive to the atomic number density $N$. Since the stability and sensitivity of many atomic and quantum technologies scale as $1/\sqrt{\mathcal{N}}$, where $\mathcal{N}$ is the effective number of atoms/ions participating in the light–matter interaction \cite{doi:10.1126/science.1192720, PhysRevLett.96.010401}, precise knowledge of the ensemble density is essential for accurate device design for optimized performance and quantitative modeling. Reliable determination of atomic number density is therefore critical for optimizing the performance of miniature atomic clocks, optically pumped magnetometers, photon-pair sources, nonlinear optical interfaces, and quantum memory systems, while avoiding density-dependent systematic shifts and unwanted broadening effects.\\

In alkali atoms, rubidium (Rb) is one of the most widely used platforms for MEMS-based vapor-cell applications due to its well-defined spectroscopic structure, strong optical transitions, and accessible vapor pressures near room temperature \cite{haupl2022spatially}. Naturally occurring Rb has two stable isotopes, \textsuperscript{85}Rb (72.2\%) \cite{steck2008rubidium} and \textsuperscript{87}Rb (27.8\%) \cite{steck2001rubidium}, both exhibiting distinct multi-level hyperfine structures that must be accurately modeled for the quantitative determination of the atomic number density. Estimating the atomic number density of warm alkali vapors typically relies on empirical vapor-pressure relations \cite{Alcock01071984}. The standard expression for the temperature-dependent vapor pressure $P_{\mathrm{vap}}(T)$ was first proposed by Killian~\cite{Killian1926} and later refined by Alcock, Itkin, and Horrigan \cite{Alcock01071984}. For rubidium, the vapor pressure $P_{\mathrm{vap}}(T)$ (in torr) as a function of temperature $T$ (in Kelvin) is given by,
\begin{equation}
\log_{10} P_{\mathrm{vap}}(T) = 2.881 + a - \frac{b}{T}, \label{emperical}
\end{equation}
where the constants $a$ and $b$ correspond to different physical phases: $a = 4.857, $ $ b = 4215 ~\text{(solid phase)},$ $
a = 4.312, b = 4040~\text{(liquid phase)}.$
Here, $P_{\mathrm{vap}}$ is expressed in torr; when expressed in atmospheres, the additive constant 2.881 is omitted. The corresponding atomic number density $N(T)$ is determined from the ideal-gas relation,
\begin{equation}
N(T) = \frac{P_{\mathrm{vap}}(T)}{k_{\mathrm{B}} T}, \label{number density formula}
\end{equation} 
where $k_{\mathrm{B}}$ is the Boltzmann constant and $P_{\mathrm{vap}}$ is converted to Pascals using $1~\text{torr} = 133.322~\text{Pa}$. \\

Although the equilibrium atomic number density of rubidium vapor can be estimated from well-established vapor-pressure relations (combining \autoref{emperical} and \autoref{number density formula}) using the measured cell temperature. These relations assume ideal thermodynamic equilibrium and do not account for variations arising from the fabrication and operating conditions of practical vapor cells. In particular, MEMS vapor cells may exhibit device-to-device variations due to uncertainties in alkali loading, surface adsorption and desorption processes, temperature gradients, long-term aging, and fabrication tolerances. Consequently, an experimental spectroscopic determination of the atomic number density provides an independent validation of the actual vapor density within a specific device and enables accurate characterization without relying solely on empirical vapor-pressure models. This empirical formulation remains the standard benchmark for validating experimentally inferred
alkali-vapor densities.\\

Various experimental techniques have been employed to determine the atomic number density in warm alkali vapor cells, including fluorescence spectroscopy \cite{Zhao_2015}, Faraday rotation \cite{Behnam}, and spin-exchange measurements \cite{Shang2024}. In fluorescence spectroscopy, the population of the excited state can be inferred from the intensity of the emitted light. In contrast, Faraday rotation correlates the polarization rotation of a probe beam with resonant atom density. Measurement of the atomic number density based on the Faraday rotation technique relies on the spin polarization of the atomic sample. It requires sophisticated magnetic-field isolation using mu-metal shields, as well as precision electronics for detecting spin polarization with balanced photodetectors. Thus, implementing this method of atomic number density measurement in field environments outside the laboratory is challenging. The spin-exchange method infers the atomic number density from collisional relaxation rates. Still, it is usually limited to specific regimes, e.g., high optical depth at substantially elevated temperatures, where direct absorption is impractical. Despite the utility of these approaches, Single-Pass Absorption Spectroscopy (SPAS) \cite{SPAS_ref} remains one of the most attractive, simple, and widely accessible methods for characterizing warm vapor systems, especially near room temperature \cite{siddons2008absolute}. However, while SPAS has been extensively used to investigate spectral lineshapes \cite{das2024direct} and transition strengths \cite{siddons2008absolute}, it has rarely been used for the explicit, quantitative determination of absolute atomic number density in a MEMS-based vapor cell. This limitation arises from the need for accurate baseline calibration and the complexities introduced by optical pumping, multi-level dynamics, and Doppler broadening in MEMS-based atomic vapor cells.\\

In this work, we present a systematic, quantitatively validated framework for determining the atomic number density of a dilute atomic vapor using a multi-level master equation model. Our theoretical approach efficiently incorporates the intrinsic differences between the two stable isotopes, \textsuperscript{85}Rb and \textsuperscript{87}Rb, enabling accurate and computationally tractable modeling of their spectroscopic signatures. By treating both isotopes within a unified formalism, the model not only facilitates direct comparison with experimental data but also offers insights into the role of hyperfine structure, optical pumping, and relaxation mechanisms, such as transit-time broadening and finite laser linewidth. The theoretical model is predominantly constrained by independently measured experimental parameters, including the cell length, laser power, beam diameter, and temperature, while the atomic number density is treated as the principal inferred quantity obtained by fitting the simulated absorption spectra with experimental absorption spectra. Although quantitative single-pass absorption spectroscopy has been extensively developed for conventional alkali vapor cells, its application to MEMS-scale (2 mm) vapor cells has remained comparatively less explored because conventional opacity-based zero-transmission calibration becomes impractical at such short optical path lengths. In this work, we employ a dark-current-based baseline normalization method that is well suited for low-optical-depth MEMS vapor cells and combine it with a unified multi-level master-equation model for quantitative atomic number density determination. The validity of the extracted number density is further established through internal cross-validation using two widely separated optical transitions (780 nm and 420 nm), as well as different vapor-cell lengths, temperatures, and probe power.\\

We demonstrate this method across markedly different optical path lengths, using both a MEMS vapor cell (2 mm, designed and fabricated in the LEOS (Laboratory for Electro-Optics Systems), ISRO (Indian Space Research Organization) \cite{giridhar2022mems}) and a commercial 100 mm (from Triad Technology Inc.) vapor cell, and at two widely separated wavelengths corresponding to the $5S_{1/2}\rightarrow5P_{3/2}$ (780.24 nm) and $5S_{1/2}\rightarrow6P_{3/2}$ (420.29 nm) transitions. This broad validation highlights the generality of the technique and its relevance for precision quantum-optics experiments, spectroscopy in weak-absorption regimes, and the characterization of MEMS-based alkali-vapor devices. Although our study focuses on Rb, the approach is readily extendable to other atomic and molecular species, making it a versatile tool for laboratory and field-deployable sensing applications.\\

The manuscript is organized as follows. In \autoref{sec:theory}, we outline the theoretical framework for modeling the absorption spectra, detailing the development of a multi-level master equation approach and integrating Doppler broadening into absorption coefficient calculations. In \autoref{sec:experiment} we describe the experimental methodology, including the layout of the optical setup and the techniques used for signal acquisition. In the following \autoref{sec:Result}, we present a comparative analysis of the simulated and experimentally observed spectra, emphasizing the influence of critical parameters, such as temperature and laser power, on the atomic number density estimation. Finally, in \autoref{sec:conclusion}, we summarize key findings and discuss potential avenues to extend this approach to other atomic and molecular systems for sensing and metrology applications.

\section{Theoretical background}\label{sec:theory}
In the presence of light–matter interaction and dissipative processes such as dephasing and spontaneous emission, the state vector alone is no longer sufficient to capture the complete dynamics of the system \cite{natarajan2015modern}.\\

Instead, the density matrix denoted by $\rho$ is used because it can describe both pure and mixed states, including decoherence, spontaneous emission, and other dissipative effects \cite{breuer2002theory, plenio1998quantum, scully1997quantum}. For an $ N$-level quantum system, the density matrix is an $ N\times N$ Hermitian matrix. In this matrix, the diagonal elements $\rho_{ii}$ indicate the probability (or population) of finding the system in the $i$-th state, and the off-diagonal elements $\rho_{ij}$ represent the quantum coherence between the states $|i\rangle$  and  $|j\rangle$. This matrix framework is essential in scenarios where a quantum system interacts with external fields or environments, leading to coherence loss. \\

\begin{figure}[h]
    \centering
    \includegraphics[width=0.9\linewidth]{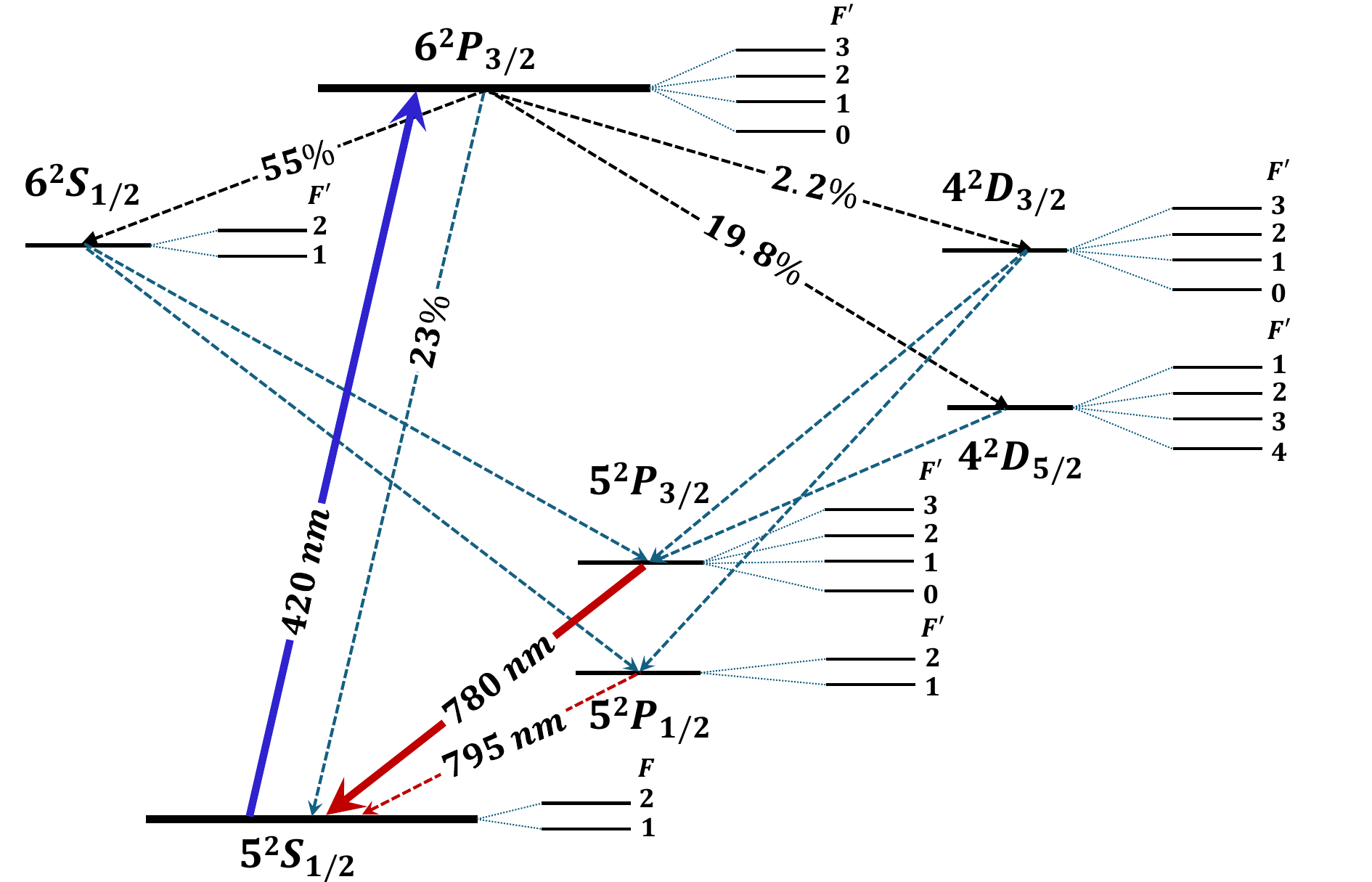}
    \caption{Energy-level diagram of rubidium ($^{87}$Rb) showing the 420 nm excitation scheme, relevant intermediate states, and decay pathways, highlighting the 780 nm and 420 nm transitions with their corresponding branching ratios \cite{noh2012transmittance} used to determine the atomic number density of the Rb atoms in the MEMS vapor cell.}
    \label{420-Rb-energy level for modeling}
\end{figure}
\begin{figure}[!h]
\begin{center}
\includegraphics[width=1\linewidth]{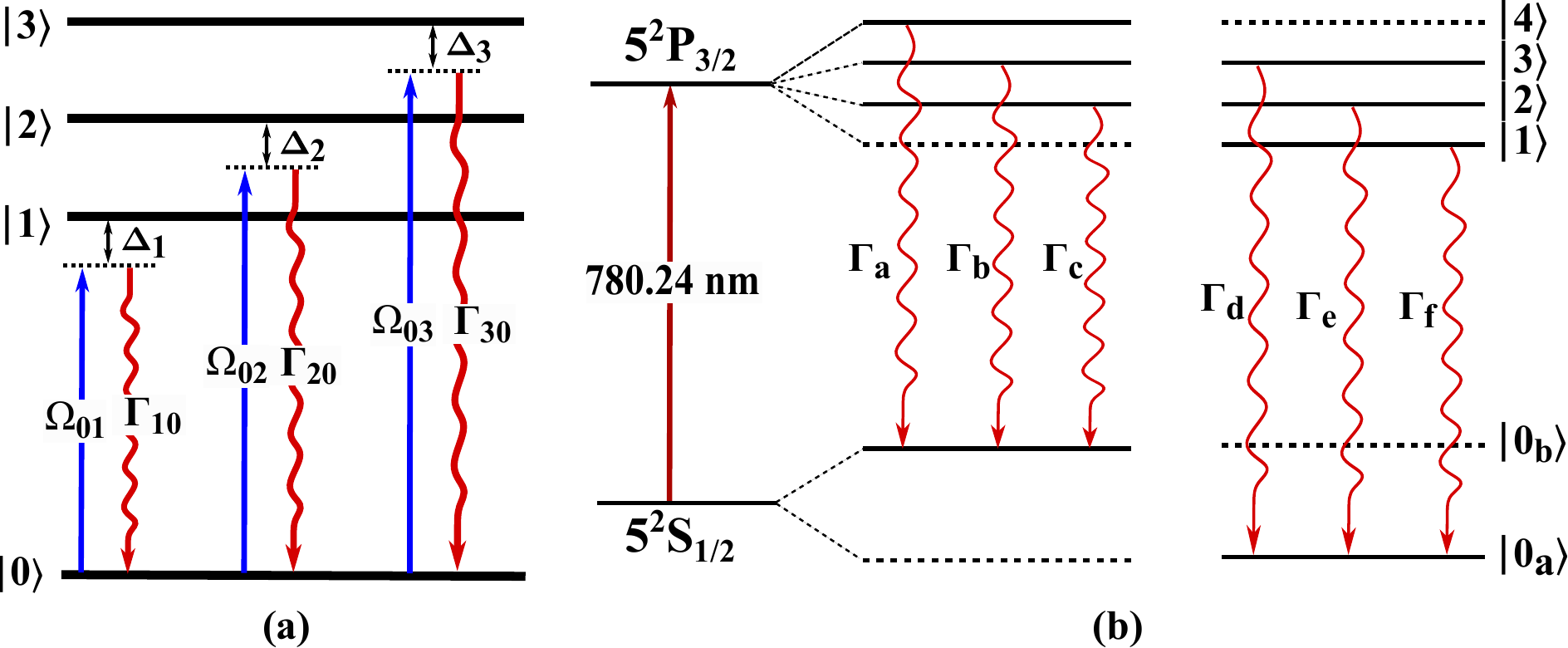}
 \caption{(a) [Left] Generalized four-level atomic system comprising a ground state  $|0\rangle$  and three excited states $|1\rangle$,  $|2\rangle$, and  $|3\rangle$. Transitions are driven by classical fields with Rabi frequencies $\Omega_{01}$ ,  $\Omega_{02}$ , and  $\Omega_{03}$ , and corresponding detunings $\Delta_1$ , $\Delta_2$ , and  $\Delta_3$ . Spontaneous emission processes occur with decay rates  $\Gamma_{10}$ , $\Gamma_{20}$, and $\Gamma_{30}$ . (b) [Right] Schematic energy level diagram representing the hyperfine structure of rubidium atoms. Transitions originate from two ground hyperfine states, $|0_a\rangle$ and $|0_b\rangle$, coupling to excited states $|1\rangle$ to $|4\rangle$ as allowed by dipole selection rules. States $|2\rangle$ and $|3\rangle$ are common to both pathways. Decay channels follow branching ratios defined by $\Gamma_b$ and $\Gamma_d$ (from $|3\rangle$) and $\Gamma_c$ and $\Gamma_e$ (from $|2\rangle$).}  \label{Rb-energy level for modeling}
 \end{center}
\end{figure}

The density-matrix formalism naturally accounts for coherence between ground and excited hyperfine states, which determines the correct steady-state population distribution and hence the absolute absorption strength. The forward model then incorporates the relevant physical mechanisms, including Doppler broadening, hyperfine optical pumping, transit-time broadening, and saturation, to predict the complete absorption spectrum. Since these effects modify both the spectral lineshape and absorption amplitude of the absorption spectra, they directly influence the quantitative determination of the atomic number density. In this work, we use the density-matrix formalism, in which the atom is treated quantum-mechanically, and the electric field is treated classically. We construct the Hamiltonian of a four-level system for the transition of Rb atom from ground state $5S_{1/2}$ to the excited state $5P_{3/2}, 6P_{3/2}$ by including the relevant atomic energy levels (as shown in \autoref{420-Rb-energy level for modeling}) and their interactions with the light field. We solve the steady state of the system using the master equation. The steady-state density-matrix solution accurately reproduces both the real (dispersion) and imaginary (absorption) components of the atomic coherence. To simulate the absorption spectroscopy of these Rb transitions ($5S_{1/2} \to 5P_{3/2}, 6P_{3/2}$), we construct a theoretical model using two separate four-level atomic systems, for each isotope: $^{85}\mathrm{Rb}$ and $^{87}\mathrm{Rb}$. Each system includes a ground state and three excited states. For a given isotope, the model distinguishes between transitions (dipole allowed) originating from the lower ($|0_a\rangle$) and upper ($|0_b\rangle$) hyperfine ground states, which couple to the common excited states $|2\rangle$ and $|3\rangle$, and to the exclusive excited states $|1\rangle$ and $|4\rangle$, respectively, as shown in \autoref{Rb-energy level for modeling}(b). This framework enables accurate modeling of optical pumping effects and decay pathways, incorporating the branching ratios between states.\\

A four-level atomic system is depicted in \autoref{Rb-energy level for modeling}(a) with one ground state $ |0\rangle$ and three excited states $\ket{1} $, $\ket{2}$, and $\ket{3}$. The total Hamiltonian of the system typically includes two principal parts: the bare Hamiltonian $H_0$ and the interaction Hamiltonian $H_I$. The bare Hamiltonian $H_0$ represents the unperturbed energy of the system, and the interaction Hamiltonian $H_I$ describes how the atom couples to the external electromagnetic field.\\

To derive such a Hamiltonian \cite{Purves2006,downes2023simple}. The interaction Hamiltonian has the general form $H_I(t) = -\textbf{{d}}\cdot \textbf{{E}(t)}$, where $\textbf{d}$ is the electric dipole operator and $\textbf{{E}(t)}$ is the electric field. Expand the dipole operator in the atomic basis and introduce the Rabi frequency $\Omega_{ij} = \frac{E_0}{\hbar}D_{ij}$ \cite{fox2006quantum}, which quantifies the strength of the coupling between the ground $|i\rangle$ and excited states $|j\rangle$ (by definition $\Omega_{ij} =\Omega_{ji}^{*}$ and $\Omega_{ii} =0$), where $E_0$ is the field amplitude and $D_{ij}$ is the dipole matrix element \cite{osti_5354747}. By applying the rotating-wave approximation (RWA) to keep only the absorption $|i\rangle \to |j\rangle$ and emission $|j\rangle \to |i\rangle$ terms, and then transforming into the rotating frame to get the time-independent Hamiltonian of the four-level system,

\begin{equation}
    H =\hbar
\begin{bmatrix}
0 & \frac{{\Omega}_{01}}{2} & \frac{\Omega_{02}}{2} & \frac{\Omega_{03}}{2} \\
\frac{\Omega_{10}^{*}}{2} & -\Delta_1 & 0 & 0 \\
\frac{\Omega_{20}^{*}}{2} & 0 & -\Delta_2 & 0 \\
\frac{\Omega_{30}^{*}}{2} & 0 & 0 & -\Delta_3
\end{bmatrix},\label{eq:hamiltonian}
\end{equation}

where $\Omega_{01}$, $\Omega_{02}$ and $\Omega_{03}$ are the Rabi frequencies associated with the three transitions, and $\Delta_1$, $\Delta_2$ and $\Delta_3$ are the corresponding detunings  \autoref{Rb-energy level for modeling}(a). The Hamiltonian in \autoref{eq:hamiltonian} represents one of the two four-level subsystems used for each isotope. The complete model consists of two such subsystems coupled through branching ratio dependent spontaneous decay within the Lindblad formalism to account for hyperfine optical pumping between the two ground hyperfine states. This constitutes an effective treatment of hyperfine optical pumping rather than an explicit six-level Hamiltonian, which is sufficient to capture the dominant physical processes relevant to the measured spectra and the quantitative determination of the atomic number density. The present model does not explicitly resolve the individual $m_F$ Zeeman sublevels and the polarization-dependent transition strengths. This approximation is justified since no external magnetic field was applied in the experiment, and the Zeeman splittings are much smaller than the Doppler-broadened linewidth. Thus, the measured absorption spectrum is an average over the unresolved Zeeman manifold, and the effect of imperfect probe polarization on the determination of atomic number density is expected to be negligible.

\subsection{Master equation for density matrix evolution}
To realistically describe the dynamics of an atomic system interacting with external laser fields and subject to environmental decoherence, one must employ an open quantum system formalism. In this context, the evolution of the system's density matrix ${\rho}$ is governed by the Lindblad master equation \cite{Fleischhauer2005, breuer2002theory}:
\begin{equation}
    \frac{\partial {\rho}}{\partial t} = -\frac{i}{\hbar} [{H}, {\rho}] + {\mathcal{L}}({\rho}),\label{eq:master_equation}
\end{equation}
where ${H}$ is the total Hamiltonian of the system, and the term ${\mathcal{L}}({\rho})$ represents the non-unitary contribution due to dissipative processes, including both decay and dephasing terms. This formulation enables the inclusion of essential effects such as spontaneous emission, dephasing, and population relaxation.\\

The general form of the Lindblad superoperator acting on the density matrix is given by \cite{breuer2002theory, Carmichael1993}:
\begin{equation} 
{\mathcal{L}}({\rho}) = \sum_k \left( {C}_k {\rho} {C}_k^\dagger - \frac{1}{2} \{ {C}_k^\dagger {C}_k, {\rho} \} \right), \label{eq:superoperator}
\end{equation}
where ${C}_k$ denotes the collapse (or jump) operators that model the interaction between the system and its environment. For pure dephasing of an excited state $|e\rangle$, the collapse operator is defined as ${C}_e = \sqrt{\gamma_e} |e\rangle \langle e|$, with $\gamma_e$ representing the pure dephasing rate associated with state $|e\rangle$. In contrast, spontaneous emission from an excited state $|e_j\rangle$ to a ground state $|g_i\rangle$ is described by the collapse operator as, $ {C}_{ij} = \sqrt{\Gamma_{ij}} |g_i\rangle \langle e_j|$, where $\Gamma_{ij}$ denotes the decay rate from $|e_j\rangle$ to $|g_i\rangle$.

\subsection{Electrical susceptibility and absorption coefficient}
In the theoretical model, the interaction between the atomic ensemble and the optical fields are described by the density matrix $\rho$, and the evolution is governed by the master equation given in \autoref{eq:master_equation}. For continuous-wave excitation, the atomic system is driven into a time-independent equilibrium due to relaxation and decoherence processes such as spontaneous emission, transit-time decay, and dephasing. Consequently, the steady-state condition $\partial \rho / \partial t = 0$ \cite{Purves2006}  is sufficient to describe the dynamics of the system. The steady-state solution of the density matrix gives the coherence terms $\rho_{ij}$ corresponding to the optical transitions, from which the complex electric susceptibility of the medium can be obtained, which characterizes its polarization response to an external electromagnetic field.  For a dipole-allowed transition between states $|i\rangle \to |j\rangle$, the complex susceptibility as a function of detuning ($\Delta, T$) can be expressed as \cite{Purves2006},
\begin{equation}
\chi_{ij}(\Delta,T) = -\frac{2 N (T) D_{ji}^2}{\hbar \epsilon_{0} \Omega_{ij}}  \rho_{ij}(\Delta),
\label{eq:5}
\end{equation}

where $N$ is the atomic number density \cite{Alcock01071984}, $\Omega_{ij}$ is the Rabi frequency of the probe field describing the light–matter coupling strength, and $\rho_{ij}(\Delta)$ denotes the optical coherence obtained from the steady-state solution of the density-matrix equation (see \autoref{eq:master_equation}). The reduced electric dipole matrix element is calculated following the convention of Siddons \textit{et al.} \cite{siddons2008absolute},

\begin{equation}
D_{ji} = \langle j |D| i \rangle = \sqrt{3}  \sqrt{\frac{3 \varepsilon_{0} h \lambda^{3} \Gamma_{ji}}{8 \pi^{2}}},\label{eq7}
\end{equation}

with $\lambda$ the transition wavelength and $\Gamma_{ji}$ the effective decay rate. The latter is expressed as $\Gamma_{ji} = C_{f}^{2} \times \mathrm{\textit{BR}_{ji}} \times \Gamma$, where $\Gamma$ is the natural decay rate, $\mathrm{\textit{BR}_{ji}}$ the branching ratio of the transition $|j\rangle \to |i\rangle$, and $C_{f}$ the Clebsch–Gordan coefficient associated with the dipole coupling \cite{siddons2008absolute, DiDomenico2011}. The additional ($\sqrt{3}$) factor in \autoref{eq7} arises from the angular-momentum convention adopted for the reduced electric dipole matrix element \cite{siddons2008absolute}. In the present work, our formulation expresses the natural linewidth as a frequency, $(\Gamma=6.065~\mathrm{MHz}$ \cite{steck2008rubidium} for 780 nm and $\Gamma=1.42~\mathrm{MHz}$ \cite{noh2012transmittance} for 420 nm), rather than as an angular frequency, ($2\pi\times6.065~\mathrm{MHz}, 2\pi\times1.42~\mathrm{MHz}$ respectively). Consequently, the equivalent expression is written in terms of the Planck constant (h) instead of the reduced Planck constant ($\hbar$). The present model does not explicitly resolve the individual $m_F$ Zeeman sublevels or polarization-dependent transition strengths. Since no external magnetic field was applied in the experiment and the Zeeman splittings remain much smaller than the Doppler-broadened linewidth, this approximation is expected to have a negligible influence on the measurement of the atomic number density.\\

In our theoretical model, we explicitly consider the hyperfine structure of the rubidium atom at 780 nm and 420 nm ($5^{2}S_{1/2} \leftrightarrow 5^{2}P_{3/2}, 6^{2}P_{3/2}$) transitions, constructing two distinct four-level systems for the two naturally occurring isotopes, $^{87}$Rb and $^{85}$Rb. As shown in \autoref{Rb-energy level for modeling}(b) for $^{87}$Rb, the hyperfine levels of the ground state are $|0_a\rangle =|F=1 \rangle$ and $|0_b\rangle =|F=2\rangle$, while the hyperfine excited state ($5^{2}P_{3/2}, 6^{2}P_{3/2}$) consists of $\lvert1\rangle $ to $ \lvert 4 \rangle$, corresponding to $\lvert F' =0\rangle$ to $\lvert F'=3 \rangle$ respectively. For $^{85}$Rb, the ground state levels are $|0_a \rangle =\lvert F=2 \rangle$ and $\lvert0_b\rangle = \lvert F=3 \rangle$, with excited state levels $\lvert1\rangle $ to $ \lvert 4 \rangle$ corresponding to $\lvert F' =1\rangle$ to $\lvert F'=4 \rangle$ respectively.\\

The optical pumping process relies on spontaneous decay from the excited hyperfine states ($\ket{F'}$) back to the ground states ($\ket{F}$). The decay pathways are governed by selection rules ($\Delta F = 0, ~\pm 1$ with $F=0 \leftrightarrow F'=0$ forbidden). Not all excited states can decay to both ground states. For $^{87}$Rb:
The excited state $\ket{F'=3}$ can only decay to the ground state $\ket{F=2}$. The excited state $\ket{F'=0}$ can only decay to the ground state $\ket{F=1}$. The intermediate excited states, $\ket{F'=1}$ and $\ket{F'=2}$, can decay to {both} ground states, $\ket{F=1}$ and $\ket{F=2}$. The effect of the optical pumping in the measurement of the atomic number density is shown in \autoref{appendix:optical_pumping}.
The relative probability of an excited atom decaying into a specific ground state is determined by the {branching ratio} of that transition. The branching ratio for a decay from an initial state $\ket{i}$ to a final state $\ket{f}$ is defined as \cite{himsworth2010rubidium, singh2015hanle},
\begin{equation}
    \text{Branching Ratio} = BR_{if}= \frac{C_f{i \rightarrow f}}{\sum_{i'} C_f{i' \rightarrow f}}\label{eq:branching},
\end{equation}
where $C_f{i \rightarrow f}$ is the transition strength from the initial state $\lvert i \rangle$ to a particular final state $\lvert f \rangle$.  Since the total electronic angular momentum $J$, nuclear spin $I$, and orbital angular momentum $L$ are identical for the $5P_{3/2}$ and $6P_{3/2}$ states, the hyperfine structure, Clebsch--Gordan coefficients, and branching ratios for individual hyperfine transitions remain unchanged relative to the $D_2$ line. Despite these similarities, the overall absorption strength of the $5S_{1/2} \rightarrow 6P_{3/2}$ transition is significantly weaker than that of the $D_2$ transition. Unlike the $D_2$ line, where spontaneous decay from the excited state occurs exclusively to the ground state, the $6P_{3/2}$ state exhibits multiple radiative decay channels, as shown in \autoref{420-Rb-energy level for modeling}. Approximately $23\%$ of the total spontaneous decay from $6P_{3/2}$ returns directly to the ground state $5S_{1/2}$ \cite{noh2012transmittance}, with the remaining population decaying via intermediate states. The reduced branching ratio to the ground state and the smaller dipole matrix element result in substantially weaker absorption on the $420~\mathrm{nm}$ transition. The branching ratios corresponding to all allowed transitions of both $^{85}$Rb and $^{87}$Rb are presented in \autoref{branching ratio}.

\begin{table}[!h]
    \centering
    \renewcommand{\arraystretch}{2}
    \setlength{\arrayrulewidth}{1pt}
        \begin{tabular}{|c|c| c| c| c|}
       
        \hline
        \multirow{2}{*}{$F_g$} & \multicolumn{4}{c |}{$F_e$} \\ 
        \cline{2-5}
        & 1 & 2 & 3 & 4 \\ 
        \hline
        2 & 1 & $\ \frac{7}{9} \ $ & $\frac{4}{9}$ & NA \\ 
        
        \ 3 \ & NA & $\frac{2}{9}$ & $\frac{5}{9}$ & 1 \\ 
        \hline

         \multicolumn{5}{c}{\textbf{(a)}} 
    \end{tabular}
    \hspace{0.5cm}
    \begin{tabular}{|c|c |c |c |c|}
        
        \hline
        \multirow{2}{*}{$F_g$} & \multicolumn{4}{c|}{$F_e$} \\ 
        \cline{2-5}
        & 0 & \ 1 \ & \ 2 \ & 3 \\ 
        \hline
        1 & 1 & $\frac{1}{2}$ & $\frac{1}{6}$ & NA \\ 
        2 & NA & $\frac{1}{2}$ & $\frac{5}{6}$ & 1 \\ 
       
        \hline
         \multicolumn{5}{c}{\textbf{(b)}} 
    \end{tabular}
    \caption{Branching ratio for the $D_2$ line of (a) $^{85}$Rb and (b) $^{87}$Rb, where 'NA' is not allowed as per the selection rule.}
    \label{branching ratio}
\end{table}
The absorption coefficient follows from the imaginary part of the susceptibility,
\begin{equation}
\alpha_{ij}(\Delta,T) = k * \mathrm{Im}[{\chi_{ij}(\Delta,T)}],
\label{eq:absorption}
\end{equation}

where $k$ is the probe wave vector, and $\chi_{ij}(\Delta,T)$ is the complex susceptibility from \autoref{eq:5}. In the four-level model considered here (see \autoref{Rb-energy level for modeling}(a)), the ground state couples to three excited states, giving rise to three distinct absorption channels $\alpha_{0j}$ ($j=1,2,3$). 

\subsection{Transit time and Doppler broadening}
The linewidth of an absorption spectrum in an atomic vapor cell is influenced by several homogeneous and inhomogeneous broadening mechanisms, among which transit-time relaxation and Doppler broadening play significant roles. The transit-time relaxation arises from the interaction time between moving atoms and the incident laser beam. As atoms with velocities from the Maxwell-Boltzmann distribution traverse the probe region, they are effectively reset upon entering and exiting. When an atom leaves the interaction volume, it is statistically replaced by a new, unpumped atom from the surrounding ensemble. From the system's perspective, this process acts as an additional relaxation pathway that contributes to linewidth broadening. Unlike spontaneous emission, which occurs only from excited states to the ground state, transit-time relaxation introduces a dephasing mechanism that affects both ground and excited states uniformly. It destroys the phase coherence of the atomic ensemble, thereby broadening the spectral line. The transit relaxation rate, denoted as $\gamma_{t}$ expressed as,
\begin{equation}
\gamma_t = \frac{\langle v \rangle }{D} \label{eq:transit},
\end{equation}
where $\langle v \rangle = \sqrt{\frac{8 k_{B} T}{\pi m}}$ is the mean thermal speed of atoms in the Maxwell–Boltzmann distribution, 
$k_{B}$ is the Boltzmann constant, $T$ is the absolute temperature of the vapor, $m$ is the atomic mass, and $D$ is the FWHM of the probe beam \cite{2022JPhB}. $D$ is taken as the geometric mean of the measured $1/e^2$ beam diameters along the major and minor axes \autoref{laser_beam_analysis}, $D_{1/e^2} = \sqrt{D_x D_y} = \sqrt{(2.42~\mathrm{mm})(1.55~\mathrm{mm})} \approx 1.94~\mathrm{mm}$, converted to an equivalent FWHM via $D_{\mathrm{FWHM}} \approx 0.589\,D_{1/e^2} \approx 1.14~\mathrm{mm}$. For the measured beam diameter ($\sim$1.94~mm), the atomic transit time through the laser beam is of the order of several microseconds, whereas the excited-state lifetime of the Rb D$_2$ transition is approximately 26~ns \cite{steck2001rubidium}. Since the transit time is more than two orders of magnitude longer than the internal relaxation time, the atomic system reaches steady state well before leaving the interaction region. A smaller beam diameter or a higher atomic temperature leads to a larger $\gamma_{t}$, corresponding to a stronger transit-induced broadening. The transit time broadening contributes an additional dephasing rate \cite{griesser2019spectral} and is added to the dephasing part of the Lindblad superoperator \autoref{eq:superoperator}.\\

This expression \autoref{eq:transit} is a first-order approximation that treats all atoms as experiencing a single effective interaction $D$, it does not account for the distribution of transit times associated with different impact parameters across the Gaussian intensity profile of the probe beam.\\

In a warm vapor, the atomic motion gives rise to Doppler broadening, an inhomogeneous effect arising from the Maxwell–Boltzmann distribution of atomic velocities. At finite temperatures, atoms move with varying velocities according to the Maxwell–Boltzmann distribution, causing each atom to experience a different effective laser frequency due to the Doppler shift. As the temperature increases, the distribution widens, resulting in a broader spectral line. The absorption coefficient derived in \autoref{eq:absorption} describes the natural line shape of an atomic transition, which follows a Lorentzian profile characterized by a homogeneous linewidth that includes contributions from spontaneous emission and transit-time relaxation (see \autoref{eq:transit}). Considering motion only along the laser propagation direction (z-axis), the Doppler-shifted detuning between the laser frequency and the atomic resonance is $\Delta_{\text{modified}} = \Delta \mp k v_z$, where the sign depends on the direction of motion of the atoms, $k$ is the wavevector, and $v_z$ is the velocity component along the $z$-axis. The overall observed line shape is obtained by convolving the Lorentzian profile with the Maxwell–Boltzmann velocity distribution $M(v)$, which accounts for the spread in velocity along the propagation direction. This convolution yields the Voigt profile, a line shape that accurately represents the combined effects of homogeneous (natural and transit-time) and inhomogeneous (Doppler) broadening mechanisms in thermal atomic vapors.\\

To obtain the Doppler-broadened spectra, the absorption coefficient from \autoref{eq:absorption} is integrated over all atomic velocities, 
 \begin{equation}
     \alpha^{\prime}_{ij}(\Delta,T) = \int_{-\infty}^{+\infty} \alpha_{ij}(\Delta - k v,T)M(v)dv. \label{eq:8}
 \end{equation}
Ideally, the simulation of the Doppler-broadened absorption coefficient (in \autoref{eq:8}) should span a velocity range that captures nearly all atoms in the vapor. However, to balance efficiency and computation time, we consider the velocity range from \(-4\sigma_v\) to \(4\sigma_v\), which includes approximately $99.99\%$ of the atomic population. Here, $\sigma_v = \sqrt{\frac{k_B T}{m}}$ denotes the root-mean-square (rms) velocity of the atoms along the laser propagation direction, where $k_B$ is the Boltzmann constant, $T$ is the temperature of the atomic vapor, and $m$ is the atomic mass.

\subsection{Light propagation through the vapor cell}

The propagation of resonant light through an atomic vapor is governed by absorption and dispersion processes that modify the transmitted intensity along the direction of propagation. For a monochromatic beam incident on a homogeneous vapor of length $z$, the transmitted intensity follows the Beer-Lambert law,
\begin{equation}
    I(z) = I_0 e^{-\alpha^{\prime}_{ij}(\Delta,T) * z},\label{eq:12}
\end{equation}
where $I_0$ is the incident intensity, and $\alpha^{\prime}_{ij}(\Delta, T)$ is the Doppler-broadened absorption coefficient (from \autoref{eq:8}), which depends on the detuning $\Delta$ of laser and the vapor temperature $T$. The absorption coefficient includes the combined effects of the atomic line profile, natural, transit time, and Doppler broadening.\\

Spectroscopic measurements are performed by monitoring the fraction of light transmitted through the vapor cell, and transmission provides a dimensionless and experimentally observable quantity that reflects the frequency-dependent absorption of the medium. The normalized transmission through the vapor cell is from \autoref{eq:12} as,
\begin{equation}
    \mathcal{T}(z) = \frac{I(z)}{I_0} = e^{-\alpha^{\prime}_{ij}(\Delta,T)*z}.
    \label{eq:beer}
\end{equation}

The present model includes hyperfine optical pumping through branching-ratio-dependent spontaneous decay between the hyperfine manifolds. The individual $m_f$ Zeeman sublevels are not explicitly included. This approximation is appropriate for the present measurements, where the Zeeman structure remains unresolved, and the measured transmission corresponds to an average over the magnetic sublevels.
\section{Experimental setup \label{sec:experiment}} 

\begin{figure}[!h]
\begin{center}
  \includegraphics[width=1\linewidth]{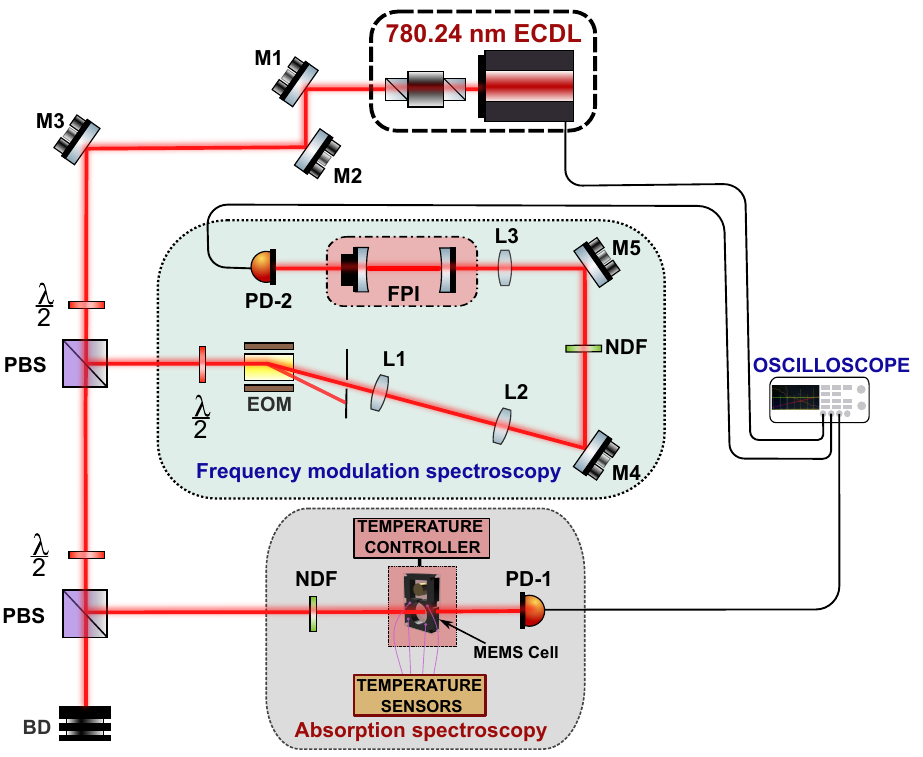}

\caption{Schematic of the experimental setup for absolute absorption spectroscopy and frequency calibration. The absorption signal transmitted through the rubidium vapor cell is detected using photodetector (PD-1), while the reference transmission signal from the Fabry–Pérot interferometer (FPI) is monitored by photodetector (PD-2). Both signals are simultaneously recorded with a digital storage oscilloscope (DSO). The piezoelectric transducer (PZT) scan signal from the external cavity diode laser (ECDL) controller is used as the trigger input to synchronize the oscilloscope trace with the frequency scan. Optical components include mirrors (M1–M5), neutral density filters (NDF), lenses (L1–L3), polarizing beam splitters (PBS), an electro-optic modulator (EOM), and a beam dump (BD). The vapor cell temperature is stabilized using a temperature controller and resistive foil heaters, which are monitored by four NTC (Negative Temperature Coefficient) temperature sensors.} \label{Experimental schematics}
\end{center}
\end{figure}

We performed SPAS \cite{SPAS_ref, siddons2008absolute} on rubidium vapor contained in both a commercial 100 mm vapor cell and a microfabricated MEMS vapor cell with an optical path length of 2 mm. The MEMS rubidium vapor cell used in this work was designed and fabricated at the Laboratory for Electro-Optic Systems (LEOS), Indian Space Research Organisation (ISRO) \cite{giridhar2022mems}. The cell contains naturally abundant rubidium vapor $(72.2\%~^{85}$Rb and $27.8\%~ ^{87}$Rb) without any buffer gas or anti-relaxation coating, and its optical windows are fabricated from Pyrex glass. Consequently, pressure broadening due to buffer-gas collisions is absent, and the measured spectra are governed by the broadening mechanisms explicitly included in our theoretical model. The measurements were carried out primarily on the $D_2$ transition of rubidium at 780.24 nm. A schematic of the experimental setup is presented in \autoref{Experimental schematics}.\\

A tunable external-cavity diode laser (ECDL, Toptica DL PRO 780) operating at $780.24$ nm, equipped with an inbuilt $35$ dB optical isolator, was used as the light source. The output beam was spatially aligned using three-axis mirrors (M1 and M2) and subsequently divided into a spectroscopy arm and a frequency-reference arm using a half-wave plate (HWP) and a polarizing beam splitter (PBS). The optical power in the spectroscopy arm was adjusted using an additional HWP-PBS combination and a neutral density filter (NDF).\\

To perform the experiment in the weak-probe regime and suppress power broadening and nonlinear optical pumping effects, an NDF was placed before the rubidium vapor cell.  The incident probe power attenuated to $10~\mu$W. Before the vapor cell, the laser beam's optical power was measured with a calibrated power meter, and its beam profile was measured with a CCD camera. The spatial profile of the measured laser beam is shown in Appendix \autoref{Laser beam image}. The measured beam diameter ($1/e^2$ diameter) is $2.42 \pm 0.04$ mm along the major axis and $1.55 \pm 0.03$ mm along the minor axis, with the principal axis rotated at 4.15$^\circ$. For further details, see Appendix \autoref{laser_beam_analysis}.  \\

The vapor cell was enclosed in a custom-fabricated heating assembly integrated with a temperature controller (Thorlabs TC300B) to facilitate spectroscopic measurements over a range of temperatures. However, the temperature displayed by the controller does not accurately reflect the actual temperature of the vapor cell due to thermal gradients in the heating assembly. Four precalibrated NTC temperature sensors were attached at different locations in the cell to accurately determine the cell temperature (as illustrated in \autoref{Experimental schematics}). The temperature readings from these sensors were recorded using an Arduino Uno, and the average value was considered as the effective vapor cell temperature for subsequent simulations. The absorption signal was detected with an unbiased photodetector (PD-1, Thorlabs PDA36A2) and recorded on one channel of a 4-channel digital storage oscilloscope (Tektronix MSO44B). \\

On the other hand, for the frequency-reference measurement, the second beam was passed through an electro-optic modulator (Qubig PM7$\_$NIR$\_$25) driven by a $25$ MHz radio-frequency (RF) signal, which modulated the frequency of the laser beam. The modulated beam was then collimated using a lens system (L1 and L2) and directed through a mode-matching lens (L3) before entering a Fabry-Pérot interferometer (Thorlabs SA30-73 with FSR= $1500$ MHz). The transmitted signal from the interferometer, including sidebands from the 25 MHz modulation, was captured by another photodetector (PD-2) and recorded on a separate channel of the oscilloscope. This signal provides a precise reference for frequency calibration. The rubidium absorption signal and the interferometer transmission peaks were acquired simultaneously within the same oscilloscope time window, ensuring synchronized data acquisition for accurate analysis. \\

For quantitative analysis of absolute absorption, it is essential to calibrate the baseline of the detected transmission signal. The baseline offset arises from the intrinsic dark current of the photodetector. This offset voltage was measured independently by blocking the probe beam and optically isolating the detector from ambient light. The measured offset was subsequently subtracted from all recorded transmission signals and used to define the zero-transmission reference level. Another way to measure the zero level is to make the atomic vapor optically opaque by significantly increasing its temperature, so that probe light is not transmitted through the medium \cite{Pizzey_2022}. Although this technique offers a physical reference point for complete absorption, it is not universally applicable. In particular, for miniaturized vapor cells (MEMS cells) with inherently low atomic density or for investigating any weak transitions that require extreme heating to achieve complete opacity, this technique becomes difficult to implement. In such scenarios, the dark-current-based method offers a more flexible and non-invasive strategy for baseline normalization.  \\

Non-linearities in both the laser frequency scan and the output power, introduced by the piezoelectric actuator used for frequency tuning, were corrected by the methodology prescribed in \cite{Pizzey_2022}. The frequency non-linearity was removed by mapping the Fabry-Pérot interferometer (FPI) transmission peaks to their known frequency separations, and the power non-linearity was removed using the off-resonant region of the measured transmitted spectrum. The off-resonant regions were fit with a higher-order polynomial to characterize the power variation across the scan, which was subsequently used to normalize the measured absorption signal.\\

To examine the applicability of the proposed (SPAS) method over substantially different optical path lengths, the same set of measurements was also performed using a commercially available rubidium vapor cell with an optical path length of 100 mm. The corresponding experimental arrangement is provided in \autoref{appendix:100mm vapor cell} (see \autoref{Experimental schematics_100 mm}).\\

In addition to the absolute absorption measurement at 780.24 nm, absorption spectroscopy was also performed on the $5S_{1/2} \to 6P_{3/2}$ transition at 420.29 nm using the same experimental methodology as shown in 
\autoref{Experimental schematics}. The optical layout and measurement protocol remained unchanged, with wavelength-appropriate substitutions for the laser source, photodetectors, and coated optical components. All data acquisition, baseline correction, and frequency calibration procedures were identical to those employed at 780.24 nm. It enables a direct comparison of number-density estimates obtained from various transitions with different strengths and saturation intensities.

\section{Results and discussion} \label{sec:Result}
To evaluate the accuracy and the applicability of the proposed model, we systematically measured the single-pass absorption spectrum over a wide range of temperatures using both MEMS and conventional vapor cells using probe beams at 780.24 nm and 420.29 nm. All experimental parameters were precisely measured and incorporated into the theoretical model, and the model spectra were then fitted to the experimental data by treating the atomic number density as a free physical parameter to achieve quantitative agreement.\\

To validate the extracted atomic number density of the rubidium vapor, the fitted values were compared with those predicted from the well-established empirical relations reported by Alcock \textit{et al.}~\cite{Alcock01071984} over a wide temperature range (298–550~K). This comparison provides a stringent consistency check of both the experimental procedure and the theoretical modeling. In the following section, we first present results from the MEMS cell at the 780 nm transition, then compare measurements at two optical transitions (at 780 nm and 420 nm), and discuss the extracted atomic number densities for each transition. The applicability of the method to a conventional 100 mm vapor cell presented in the \autoref{appendix:100mm vapor cell} and the measurements beyond the weak-probe regime is subsequently presented in the \autoref{appnedix: high_power}. 

\subsection{Validation of the SPAS model using the MEMS vapor cell at 780 nm}

\begin{figure}[!h]
    \centering
    \includegraphics[width=1\linewidth]{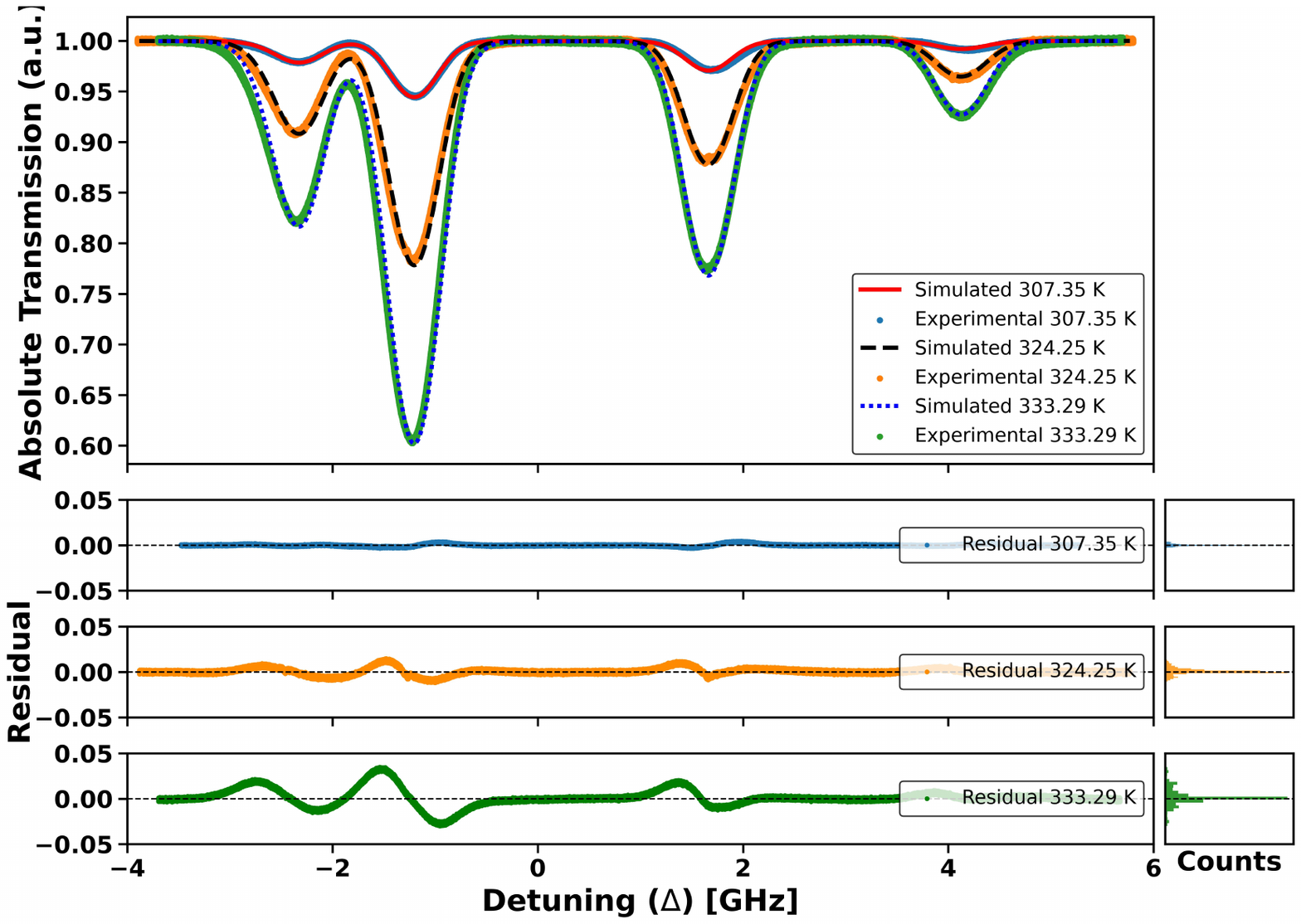}
    \caption{Comparison of the simulated (solid/dashed/dotted lines) and experimental transmission spectra of the rubidium MEMS vapor cell for the $5S_{1/2}\to 5P_{3/2}$ (780.24 nm) transition at temperatures of 307.35 K (blue), 324.25 K (orange), and 333.29 K (green). The corresponding lower panels show the residuals, defined as the difference between the experimental and simulated normalized transmission spectra, together with the residual histogram.}\label{fig:Mems_residual}
\end{figure}

For the MEMS vapor cell (length $2~$mm) containing naturally abundant rubidium (72.2\% $^{85}$Rb and 27.8\% $^{87}$Rb), single-pass absorption spectra were recorded at various temperatures: ranging from $293 - 353$K using a probe power $\sim 10~\mu$W. The corresponding experimental and theoretically fitted spectra, along with their residuals, are shown in \autoref{fig:Mems_residual}. The fitted spectra yield coefficients of determination $R^2 > 0.99$ for all temperatures, with normalized RMS residuals $0.11\%$ (at $307.35$ K), $0.34\%$ (at $324.25$ K) and $0.94\%$ (at $333.29$ K). These values confirm that the overall discrepancies between the measured and fitted spectra remain well below $1\%$ of the normalized transmission. The goodness of fit was further assessed using the reduced chi-square statistic. The fitted spectra yielded $\chi^2_{\nu}$ = 1.00, 1.03, and 1.02 for the measurements at 307.35 K, 324.25 K, and 333.29 K, respectively. Since all reduced chi-square values are close to unity, the results indicate that the spectral model provides a good description of the experimental data within the estimated measurement uncertainty. \\

To quantitatively assess the agreement between the experimental and simulated spectra, we define the residual as the point-by-point difference between the measured and simulated transmission $R(\Delta)=T_{\mathrm{exp}}(\Delta)-T_{\mathrm{sim}}(\Delta),$ where $T_{\mathrm{exp}}$ and $T_{\mathrm{sim}}$ denote the experimental and simulated normalized transmission, respectively, and $\Delta$ is the laser detuning. The residual distributions exhibit weak systematic structure at higher temperatures. This behavior is primarily attributed to the increased optical depth at higher atomic number densities, which produces deeper and steeper absorption features. In these regions, small imperfections in the frequency-axis calibration, baseline correction, and simplified line-shape model translate into larger deviations in transmission because of the increased spectral gradient. Nevertheless, the residuals remain small compared with the overall spectral variation, and the reduced chi-square values remain close to unity, indicating that the model continues to provide an accurate description of the measured spectra. Their influence on the extracted atomic number density is included in the systematic uncertainty budget. The cell temperature was monitored using precalibrated NTC temperature sensors with an intrinsic accuracy of $\pm~0.3$ K. The cell was mounted in a machined aluminum block equipped with four symmetrically placed sensors to ensure uniform heating. At elevated temperatures, a residual temperature gradient of approximately 1–2 K was observed, which accounts for the slightly larger error bars in the extracted atomic number density at higher temperatures.\\

The validation between experiment and proposed model is not limited to the short interaction length of the MEMS cell; we repeated the same measurements using a commercial 100 mm rubidium vapor cell. The corresponding spectra and detailed analysis are presented in \autoref{appendix:100mm vapor cell} (see \autoref{fig:residual_mix_cell_final_decay}), where the model reproduces the experimental observations with good agreement over the temperature range of $293.45-339.65$ K.

\subsection{Validation using the 780 nm and 420 nm transitions}
\begin{figure}[!h]
    \centering
    \includegraphics[width=1\linewidth]{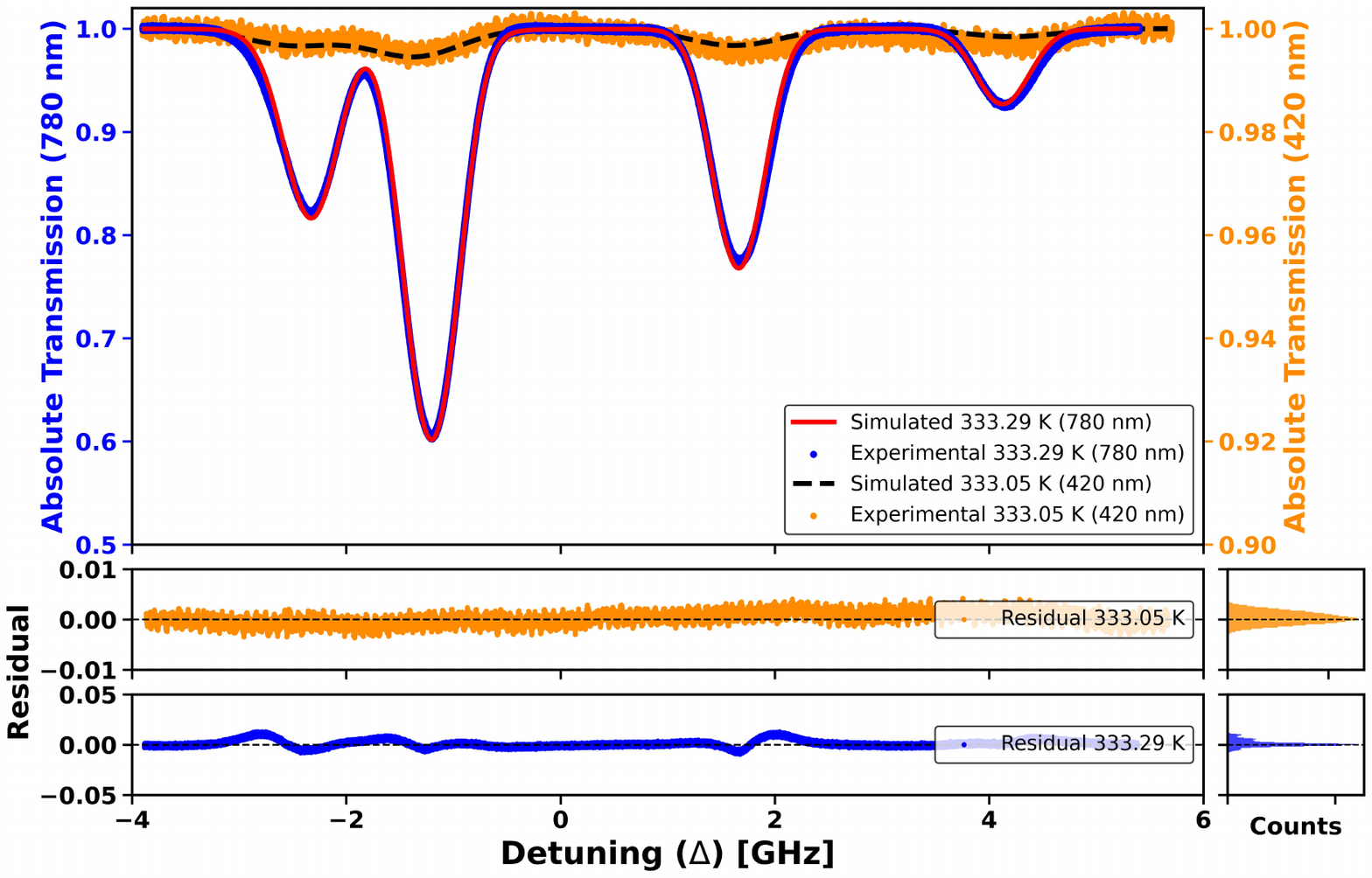}
    \caption{Experimental and simulated transmission spectra comparing the 780~nm and 420~nm rubidium transitions at the optical power of $\sim 10 ~\mu$W and $\sim 8 ~\mu$W for 780~nm and 420~nm transitions and nearly identical temperatures (333.05 K and 333.29 K). The upper panel shows simulated spectra (solid red: 780~nm, dashed black: 420~nm) with corresponding experimental data (blue and orange markers, respectively). The middle and lower panels display the residuals for the 420~nm and 780~nm fits, respectively, while the adjacent histograms illustrate the distribution of the residuals, demonstrating that the residuals remain centered around zero and providing a visual assessment of the fit quality.}
    \label{fig:strength compare}
\end{figure}

In addition to spectroscopy on the $D_2$ transition ($5S_{1/2} \rightarrow 5P_{3/2}$) at 780.24 nm, we also performed measurements using a $420~\mathrm{nm}$ laser addressing the $5S_{1/2} \rightarrow 6P_{3/2}$ transition. \autoref{fig:strength compare} compares representative transmission spectra recorded at 780 nm and 420 nm under nearly identical temperature conditions and at the normalized probe power $\sim 10 ~\mu$W and $\sim 8 ~\mu$W, respectively. Probe powers are reported in absolute units rather than normalized to the saturation intensity, since significantly different saturation-intensity values have been reported for the 420~nm transition \cite{sinha2026precision, 10.1063/1.5006962, 10.1088/1674-1056/ae194c, vernier2010enhanced}. The fitted spectra yield coefficients of determination $R^2 > 0.99$ for both wavelengths, with normalized RMS residuals of $0.35\%$ for the 780.24 nm spectra and $0.12\%$ for the 420.29 nm spectra, indicating that the model reproduces the measured transmission spectra to better than $0.4\%$ over the full spectral range. Further, the reduced chi-square values $\chi^2_{\nu}=1.12$ (780.24~nm) and $\chi^2_{\nu}=1.07$ (420.29~nm), supporting the consistency of the model with the experimental data. 
As expected, the absorption strength of the 420.29 nm transition is significantly weaker than that of the 780.24 nm ($D_2$) transitions, due to the only 23\% of the total population goes directly back to the ground state. At a cell temperature of $333.29$ K and $333.05$ K, the extracted atomic number densities are $3.36 \pm0.15\times10^{17}~\mathrm{m^{-3}}$ and $3.99 \pm 0.13\times10^{17}~\mathrm{m^{-3}}$ for the 780.24 nm and 420.29 nm transitions, respectively.\\

The weaker absorption at 420.29 nm, the corresponding signal-to-noise ratio is reduced by a factor of approximately 18.7 (25.4 dB) relative to the 780.24 nm measurements. Nevertheless, the statistical uncertainty in the extracted atomic number density remains small because the fitting procedure utilizes the entire spectral profile, including the resonance frequencies, linewidths, and relative hyperfine transition strengths, rather than relying solely on the absorption amplitude. The dominant contribution to the overall uncertainty arises from the uncertainty in the vapor-cell temperature, while the reduced signal-to-noise ratio of the 420.29 nm measurements provides a comparatively smaller contribution to the uncertainty budget. \\

Importantly, the extracted atomic number density from both 780~nm and 420~nm absorption measurements agrees to within experimental uncertainty, despite the different absorption strengths, hyperfine structures, and effective transition moments. This wavelength-independent agreement provides a stringent cross-check of the model: if $N$ extraction were sensitive to fitting artifacts, polarisation conventions, or model assumptions, the two wavelengths would yield discrepant densities. \\

The successful application of the model to MEMS cells has important implications for the development of chip-scale atomic devices, where precise knowledge of the atomic number density is crucial for optimizing sensor performance. The ability to accurately determine the vapor density in these miniaturized systems enables more accurate calibration and characterization of MEMS-based atomic, quantum, and precision timing sensors.\\

An analogous comparison was carried out using the commercial 100 mm rubidium vapor cell on the 420.29 nm transition at cell temperatures of 310.65 K and 331.75 K. The measured transmission spectra exhibit consistent agreement between the measured and simulated spectra as observed for the MEMS vapor cell. The corresponding results are provided \autoref{appendix:100mm vapor cell} (see \autoref{fig:residual plot_comparision}), further demonstrating that the proposed SPAS model remains applicable over substantially different optical path lengths.

\subsection{Temperature dependence of the extracted atomic number density}
The atomic number density is an intrinsic thermodynamic property of the rubidium vapor, determined solely by the vapor pressure at a given temperature. Consequently, for a given temperature, the extracted number density should be independent of the specific excited state used for optical interrogation, provided that the underlying spectroscopic model accurately describes the atom-light interaction. This provides a strong internal consistency check for the proposed SPAS model, as the two transitions investigated in this work differ significantly in wavelength, transition strength, and spontaneous decay rate.\\

\begin{figure}[!h]
    \centering
    \includegraphics[width=1\linewidth]{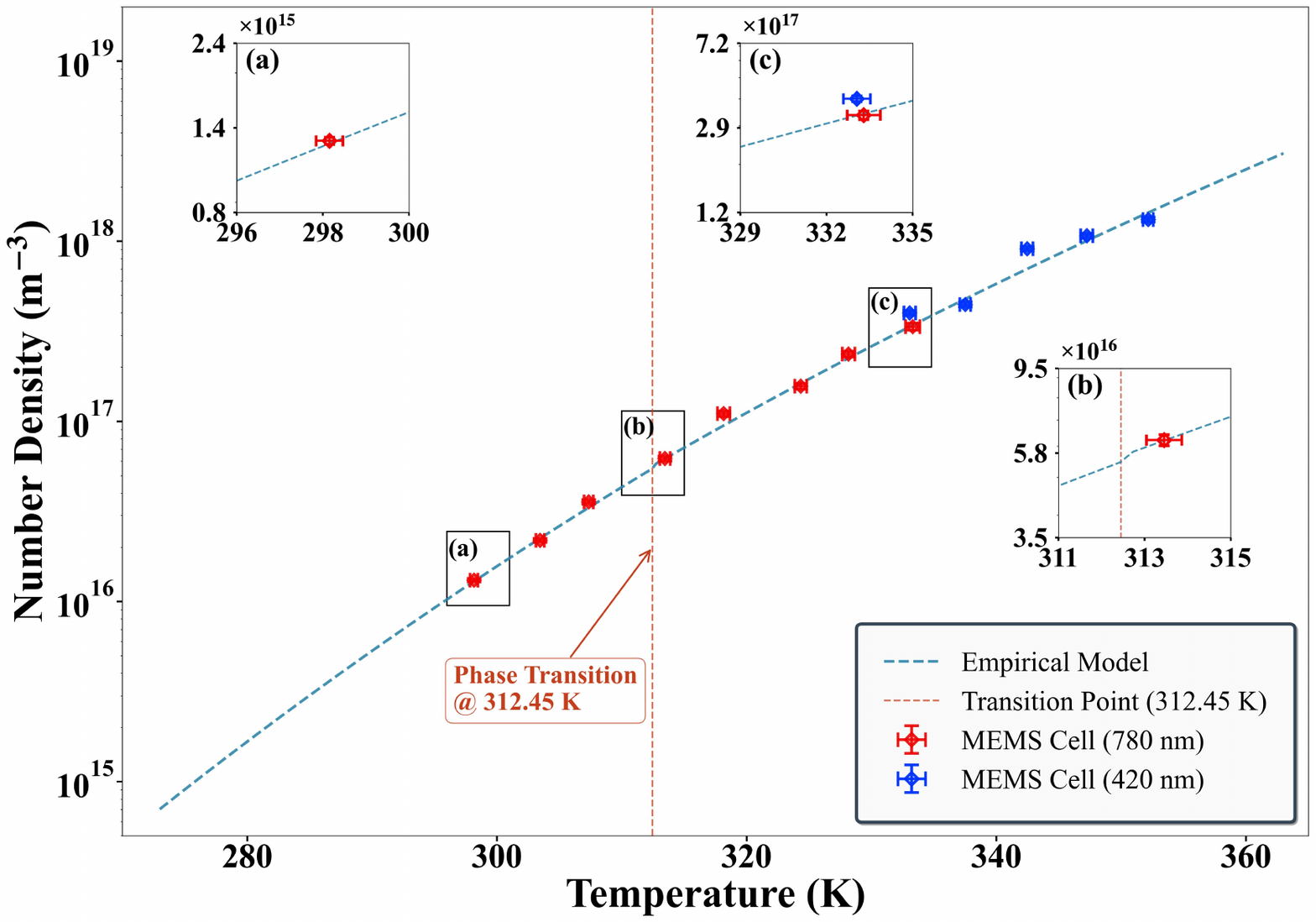}
    \caption{Atomic number density of Rb vapor as a function of temperature for the MEMS vapor cell. Experimental number densities extracted from absorption measurements at 780 nm (red diamonds) and absorption measurements at 420 nm (blue diamonds) are compared with the empirical vapor pressure model (dashed line). The vertical dashed line indicates the phase transition temperature at 312.45 K. Insets (a), (b), and (c) show magnified views around 298 K, 313 K, and 332 K, respectively. The horizontal error bars correspond to the standard error of the mean (SEM) of the temperature obtained from the five independent temperature sensor readings. The vertical error bars represent the propagated uncertainty in the extracted atomic number density, obtained from the uncertainty analysis of the SPAS model.}
    \label{fig:number_density_mems}
\end{figure}
The extracted atomic number density of rubidium in the MEMS cell as a function of temperature is shown in \autoref{fig:number_density_mems}, where results obtained from 780.24 nm ($5S_{1/2}\rightarrow5P_{3/2}$) and 420.29 nm ($5S_{1/2}\rightarrow6P_{3/2}$) absorption measurements over a range of temperatures $293 - 353$ K. The experimentally determined number densities are compared with the values predicted from the empirical rubidium vapor-pressure relation of Alcock \textit{et al.} (\autoref{emperical}). For both probe wavelengths, the extracted number densities follow the empirical vapor-pressure relation closely over the entire temperature range, a few data points exhibit deviations that are comparable to or exceed their estimated uncertainties.  Although absorption data were acquired over a broad temperature range at both probe wavelengths, only representative data points at selected temperatures are displayed in the figure to maintain visual clarity.\\

As shown in \autoref{fig:number_density_mems}, the error bars in the MEMS cell measurements are substantially smaller, indicating the improved temperature uniformity and more precise determination of the atomic number density. These observations emphasize that precise temperature monitoring and control are crucial for reducing uncertainty in vapor density measurements, particularly at higher operating temperatures.\\

We performed the same set of measurements with a commercial 100 mm Rb vapor cell with both wavelengths, 780 nm and 420 nm. Since the atomic number density is a thermodynamic property of the vapor, it is independent of the optical interaction length. The results presented in \autoref{appendix:100mm vapor cell} (see \autoref{fig:number_density_100mm}), confirm that the extracted number density is independent of the optical path length and validating the robustness of the proposed SPAS model.\\

The successful reproduction of the empirical vapor-pressure curve using both standard ($100$~mm) and MEMS ($2$~mm) vapor cells demonstrates the scalability of the absorption-based atomic number density determination method. This result is significant for precision spectroscopy and the development of compact, MEMS-based atomic devices, where accurate control of vapor density is crucial for optimal performance.

\subsection{Uncertainty analysis and sensitivity}
To quantify the uncertainty in the extracted atomic number density, a sensitivity analysis was performed by independently perturbing each experimentally measured input parameter while keeping all remaining parameters fixed. For each perturbed parameter, the absorption spectrum was refitted and the corresponding variation in the extracted number density was used to evaluate the sensitivity coefficient through the central finite-difference approximation,
\begin{equation}
\frac{\partial N}{\partial x}
\approx
\frac{N(x+\Delta x)-N(x-\Delta x)}
{2\Delta x},
\end{equation}
where $x$ denotes the experimental parameter and $\Delta x$ its associated measurement uncertainty. The uncertainty contribution from each parameter was then calculated as
\begin{equation}
\sigma_N^{(x)}
=
\left|
\frac{\partial N}{\partial x}
\right|
\sigma_x,
\end{equation}
The uncertainty budget is summarized in \autoref{tab:uncertainty} for the $T=333.29$~K and probe power $\approx10~\mu$W of MEMS cell at 780 nm.  The table lists the uncertainty contribution associated with each independently measured input parameter. The sensitivity coefficients $\partial N/\partial x$ were evaluated by independently perturbing each input parameter while keeping all remaining parameters fixed. Among the investigated parameters, the temperature uncertainty ($\pm0.3$~K) contributes less than $0.02\%$ to the extracted number density because the temperature enters only through the Doppler width ($\propto\sqrt{T}$) and transit-time broadening ($\propto\sqrt{T}$), both of which vary slowly over the uncertainty of $\pm$0.3 K of the NTC temperature sensor. The dominant systematic contributions are the systematic due to the beam-diameter (geometric mean of major and minor axis of the elliptical beam) ($0.73\%$), probe-power ($0.76\%$), and baseline normalization ($1.71\%$). Adding these in quadrature, the total systematic uncertainty is $\sim 2.01\%$.\\
\begin{widetext}
\begin{table*}[t]
\caption{Uncertainty budget for the extracted atomic number density obtained from the 780~nm transition at $T=333.29$~K using a 10~$\mu$W probe beam.}
\label{tab:uncertainty}
\centering
\renewcommand{\arraystretch}{1.3}
\setlength{\tabcolsep}{8pt}

\begin{tabular}{lcccc}
\hline\hline
\textbf{Parameter} &
\textbf{Nominal Value} &
\textbf{Uncertainty} &
\boldmath$\sigma_N$ (\textbf{m}$^{-3}$) &
\textbf{Relative Uncertainty (\%)}\\
\hline
Temperature &
333.29 K &
$\pm0.30$ K &
$4.76\times10^{13}$ &
0.014 \\

Beam diameter &
0.485 mm &
$\pm0.006$ mm &
$2.47\times10^{15}$ &
0.734 \\

Laser power &
10 $\mu$W &
$\pm2.5\%$ &
$2.57\times10^{15}$ &
0.764 \\

Baseline normalization &
16.1 mV &
$\pm0.875$ mV &
$5.76\times10^{15}$ &
1.71 \\
\hline
\textbf{Total systematic uncertainty} &
&
&
$\sigma_{\rm sys}$ &
\textbf{2.01} \\
\hline\hline
\end{tabular}
\end{table*}
\end{widetext}

The statistical uncertainty is evaluated from the standard error of the mean (SEM) of the four independently measured temperature sensors and propagated through the density-extraction procedure. In both the \autoref{fig:number_density_mems} and \autoref{fig:number_density_100mm}, the plotted error bars represent the propagated statistical uncertainty with the systematic uncertainty.\\

The contribution of the frequency calibration was also evaluated by perturbing the calibrated frequency scale by $\pm1$~MHz, which is the linewidth of the reference peaks of the Fabry-Pérot interferometer. The difference in the extracted atomic number density was negligible. This negligible contribution is expected because the calibration uncertainty is more than two orders of magnitude smaller than the Doppler-broadened linewidth and primarily introduces a small horizontal shift of the spectrum without affecting the absorption amplitude.\\

All input parameters except the atomic number density $N$ are independently measured or constrained by literature values prior to fitting: the temperature $T$ is measured by four NTC sensors ($\pm 0.3$~K), the beam diameter $D$ is characterized by a CCD imaging array to ISO~11146 standard, the probe power $P$ is measured by a calibrated power meter, the baseline offset is determined from the photodetector dark-current measurement and verified against the off-resonant region of the spectrum, the frequency axis is calibrated using the FPI reference peaks, and isotopic abundances, natural linewidths, and branching ratios are fixed at literature values
\cite{steck2001rubidium, steck2008rubidium}.  Consequently, the atomic number density remains the principal inferred parameter.

\subsection{Performance of SPAS model beyond the weak-probe regime} \label{strong-probe}
The theoretical model developed in this work is primarily intended for quantitative absorption spectroscopy in the weak-probe regime  \cite{weakprobe}, where the probe intensity is sufficiently lower the saturation intensity.
To verify the validation of our model beyond the weak-probe regime, we perform the absorption spectroscopy measurement at a higher probe power of $100~\mu\mathrm{W}$. At these higher powers, saturation effects become significant and the absorption is no longer accurately described by the linear Beer–Lambert approximation. \\

\autoref{fig:powerdependence} compares the measured transmission spectra with simulations obtained using the iterative susceptibility model. The calculated spectra continue to reproduce the measured spectral profile and absorption depth with good agreement with $R^2 > 0.99$, confirming its robustness beyond the weak-probe regime.

Although the difference between the conventional Beer-Lambert approach and the iterative model is relatively small for the 100 mm vapor cell used in this work, the iterative treatment becomes increasingly important for longer interaction lengths or higher optical depths, where the probe intensity changes appreciably along the propagation direction. Representative experimental results and detailed descriptions of the iterative propagation model are presented in \autoref{appnedix: high_power}, with the complete analysis is provided therein.\\

\section{Conclusion}\label{sec:conclusion}

We have presented a quantitative method for extracting the absolute atomic number density of warm rubidium vapor in MEMS vapor cells using SPAS. The approach is based on a multi-level Lindblad master equation formalism that incorporates hyperfine structure, Doppler averaging, optical pumping, and transit-time relaxation. All experimentally measured parameters, including vapor-cell temperature, optical path length, beam diameter, and optical power, are directly included in the model.\\

The method was experimentally validated using a MEMS vapor cell over a broad temperature range. Despite the intrinsically low optical depth associated with such short path lengths, the extracted number densities exhibit agreement with the empirical vapor-pressure relation of Alcock \textit{et al.}, while providing direct experimental validation of the vapor density in the investigated device. Consistent density values were obtained from absorption measurements on both the strong $5S_{1/2}\!\rightarrow\!5P_{3/2}$ (780.24~nm) transition and the weak $5S_{1/2}\!\rightarrow\!6P_{3/2}$ (420.29~nm) transition shows the applicability of the technique in weak-absorption regimes relevant to MEMS-scale devices.\\

The intrinsic photodetector dark current was used as a zero-transmission reference to provide a stable, noninvasive calibration method well suited to MEMS vapor cells, where complete absorption cannot be achieved without impractically high temperatures. The resulting fits reproduce the measured spectra with high fidelity, yielding $R^2 > 0.99$, normalized RMS residuals below $1\%$, and reduced chi-square values close to unity, confirming the robustness and accuracy of the proposed model for extracting the atomic number density. \\

The theoretical model described in this work may be extended to predict absorption spectra and to calculate the atomic number density of other atomic or molecular vapors, provided that the energy-level diagrams are known. Future work could also include additional transitions in the Rb atom to establish a complete formalism for the saturation absorption spectroscopy. These extensions will enhance the model's versatility in atomic physics applications and the development of quantum technology.

\subsection*{Author Contributions}
Sumit Achar and Shivam Sinha contributed equally to this work as co-first authors, theoretical model development, experimental design, data analysis, and manuscript writing. Ezhilarasan M. and Chandankumar R.: experimentation, data collection and analysis support. Arijit Sharma: conceptualization of the project, manuscript writing, review, editing, and overall project supervision.

\subsection*{Acknowledgments}

Sumit Achar gratefully acknowledges financial support from the Council of Scientific \& Industrial Research (CSIR, Govt. of India) through a Senior Research Fellowship (SRF). Shivam Sinha gratefully acknowledges the financial assistance provided by IIT Tirupati, facilitated through the half-time research assistantship (HTRA). Arijit Sharma thanks Dr. M. S. Giridhar, Dr. Jiju John and Dr. S. P. Karanth from the LEOS-ISRO (Laboratory for Electro-Optic Systems - Indian Space Research Organisation), Bengaluru, for sharing a sample MEMS rubidium (Rb) vapor cell for spectroscopic quantification through the ISRO RESPOND project.

\section*{Funding}
The present work is supported by the financial support from IIT Tirupati through the CAMOST grant and ISRO RESPOND project number RES-URSC-2022-005.

\subsection*{Conflicts of Interest}

The authors declare no conflicts of interest.

\section*{Data availability}
The data that support the findings of this study are available upon reasonable request from the authors.
\bibliography{reference, review}



\appendix
\renewcommand{\thefigure}{A\arabic{figure}}
\setcounter{figure}{0}

\section{Measurement of laser beam diameter} \label{laser_beam_analysis}
\begin{figure}[!h]
    \centering
    \includegraphics[width=1\linewidth]{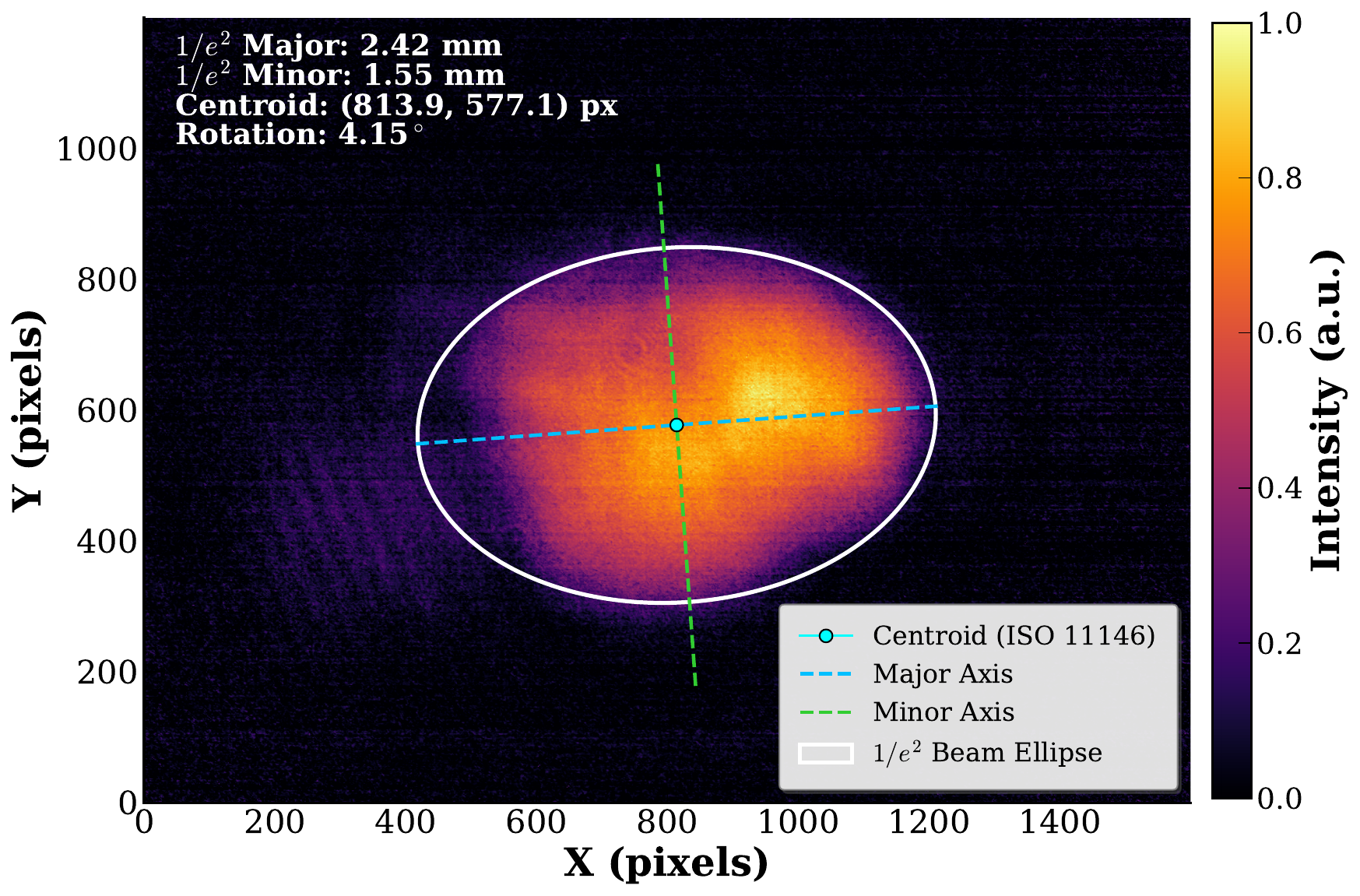}
    \caption{Beam profile analysis of the laser used in the experiment, measured using a CCD camera. The green and blue line shows the minor axis and major axis, respectively; the blue dot represents the centroid, and the white illustrates the $1/e^2$ elliptical beam diameter of the laser beam. This characterization is consistent with a distorted (elliptical) Gaussian beam and follows the ISO 11146 standard \cite{ISO11146-1_2021} for laser beam profiling.}
    \label{Laser beam image}
\end{figure}

The Rabi frequency plays a crucial role in determining the absolute absorption profile of alkali atoms. To accurately calculate the Rabi frequency, we need to know the beam diameter, as the Rabi frequency inversely depends on the beam’s cross-sectional area. Thus, a precise measurement of the laser beam diameter is very crucial. As the spatial profile of our laser beam is elliptical, the best way to measure the beam diameter ($1/e^2$ diameter) is using the centroid method.  To accurately determine the beam dimensions, we captured its image using a CCD camera (with a pixel size of 2.8 $\mu$m) at a position just before the vapor cell, as illustrated in \autoref{Laser beam image}.  
The recorded beam profile was analyzed according to the ISO 11146 standard \cite{ISO11146-1_2021}, which prescribes the use of second-order statistical moments to determine the beam width and orientation. Specifically, first-order moments were used to identify the centroid of the beam (center of mass), while second-order moments provided information about the spatial variance and the azimuthal tilt of the beam \cite{MONTERO2024101830}. The beam diameters corresponding to the $1/e^2$ intensity level were extracted along the principal axes of the elliptical profile, which were rotated with respect to the CCD pixel grid. These diameters, initially calculated in pixel units, were converted to physical dimensions using the known pixel size. Based on this analysis, the beam diameter along the major axis was measured to be $2.42~\pm~0.04$ mm, while the minor axis diameter was found to be $1.55~\pm~0.03$ mm with the rotation angle of the principal axis 4.15$^\circ$. 
\section{Characterization of atomic number density for 100 mm Rb vapor cell } \label{appendix:100mm vapor cell}
\begin{figure}[h!]
\begin{center}
\includegraphics[width=1\linewidth]{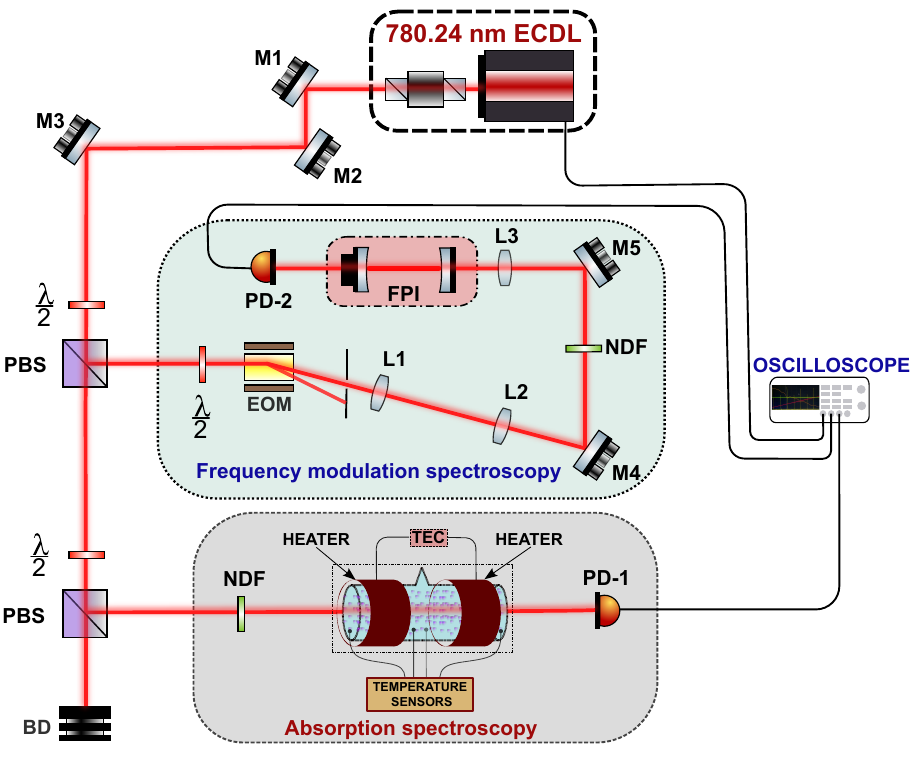}
\caption{Schematic of the experimental setup for absolute absorption spectroscopy and frequency calibration for a 100 mm long vapor cell.} \label{Experimental schematics_100 mm}
\end{center}
\end{figure}
We also determine the atomic number density of Rb vapor in a 100~mm long vapor cell using absorption spectroscopy and quantitative spectral fitting. This analysis establishes a reliable, self-consistent characterization of the atomic number density in the 100~mm cell. The experimental schematic for the SPAS is shown in \autoref{Experimental schematics_100 mm}.\\

\begin{figure}[!h]
    \centering
    \includegraphics[width=1\linewidth]{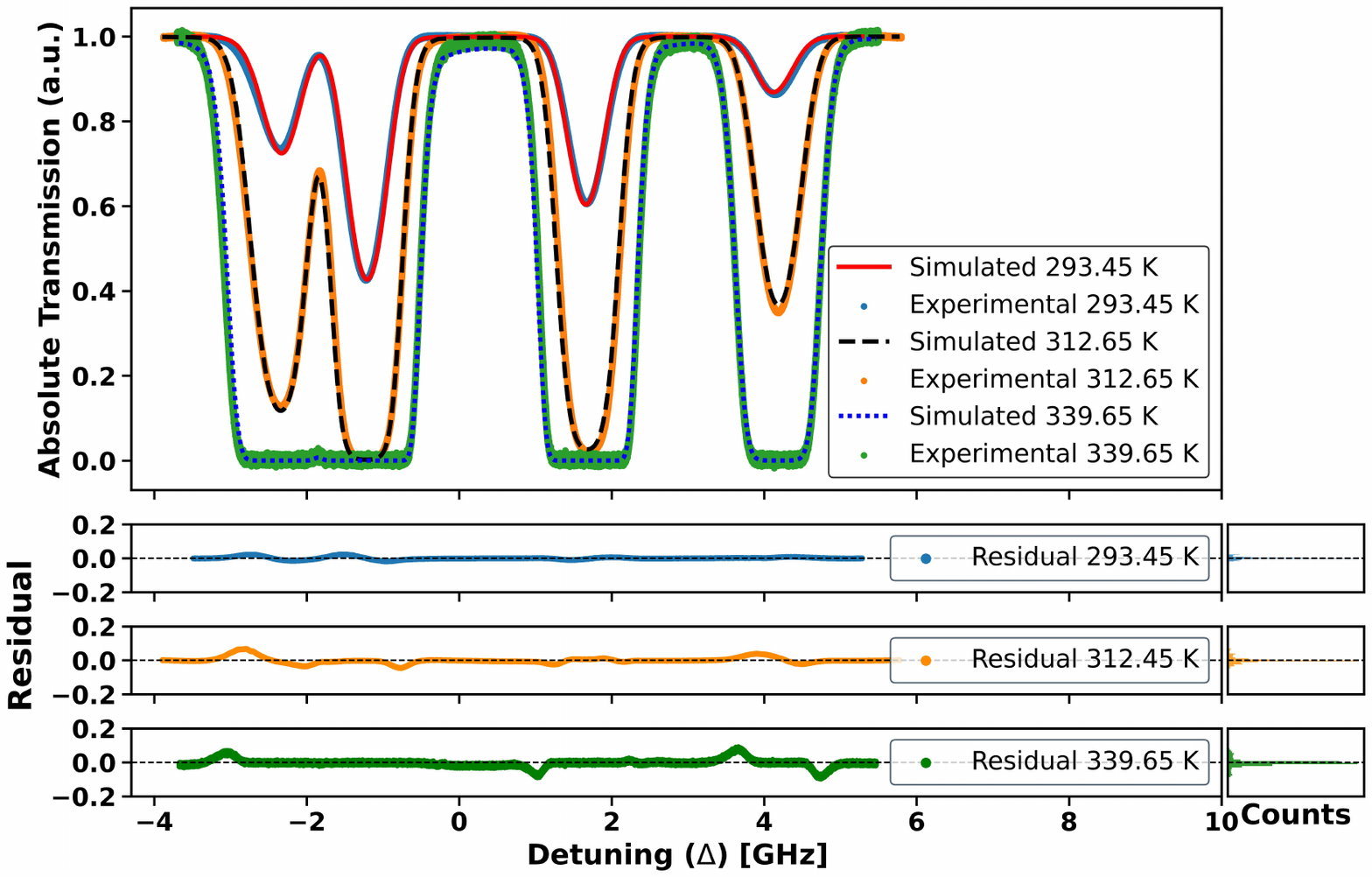}
    \caption{Experimental and fitted absorption spectra as a function of detuning for three different temperatures: 293.45 K (red solid line for fit, blue dots for experiment), 312.65 K (black dashed line for fit, orange dots for experiment), and 339.65 K (blue dotted line for fit, green dots for experiment) for the 780~nm transition with a 10~$\mu$W probe beam. The lower panels show the corresponding residuals (experimental minus fitted). The fitted spectra exhibit excellent agreement with experimental data across all temperatures, with residuals remaining within $\pm~0.25$ in absolute transmission.}
    \label{fig:residual_mix_cell_final_decay}
\end{figure}

\autoref{fig:residual_mix_cell_final_decay} shows the measured absorption spectra as a function of laser detuning at three cell temperatures: $293.45~ \pm ~0.3$ K, $312.65 ~\pm~ 0.95$ K, and $339.65 ~\pm~ 2.5$ K. We fit the experimental spectra using a model that includes Doppler broadening, natural linewidth, and temperature-dependent atomic number density. The fitted spectra accurately reproduce the peak absorption, relative transition strengths, and off-resonant wings at all temperatures. The residuals shown in the lower panels of \autoref{fig:residual_mix_cell_final_decay}, remain within $\pm 0.25$ in absolute transmission.\\

\begin{figure}[!h]
    \centering
    \includegraphics[width=1\linewidth]{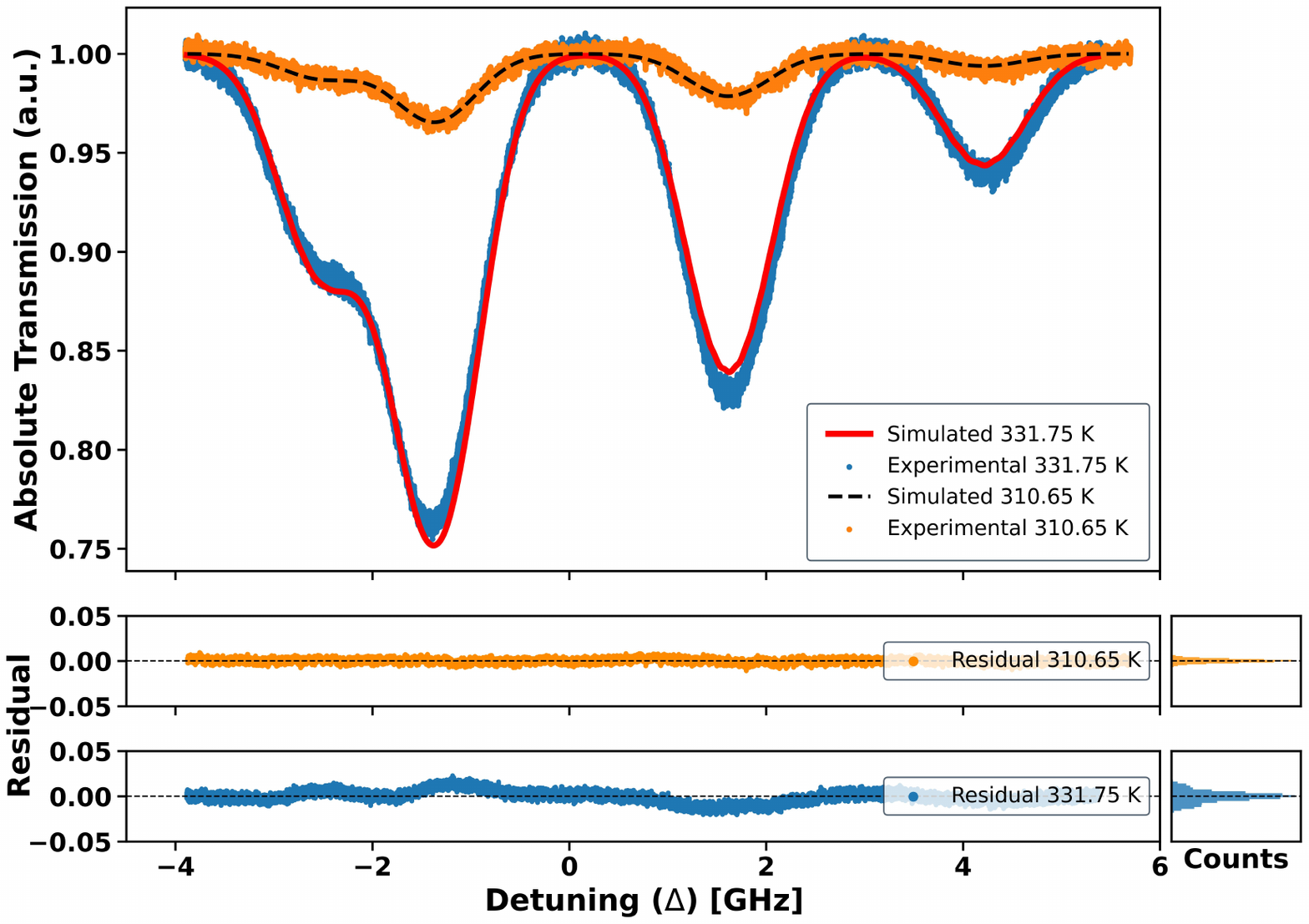}
    \caption{Comparison of simulated and experimental transmission spectra of the 100 mm vapor cell at 310.65 K and 331.75 K for the 420~nm transition with an $\sim$8~$\mu$W probe beam. Solid/dashed lines represent the simulated spectra, and blue/orange markers show the corresponding experimental data at 331.75 K and 310.65 K, respectively. The lower panels display the absolute residuals (experiment -- simulation) for each temperature along with the residual histogram.}
    \label{fig:residual plot_comparision}
\end{figure}

\autoref{fig:residual plot_comparision} presents the comparison of experimentally measured absorption spectra for $5S_{1/2}\to 6P_{3/2}$ transition and simulated spectra at 310.65 K and 331.75 K temperature using a 100 mm vapor cell. The simulated spectra reproduce both the magnitude and spectral shape of the measured transmission with good agreement. The residuals remain small across the full frequency scan, confirming that the extracted number densities provide a quantitatively accurate description of the rubidium vapor in the 100~mm cell. This analysis establishes a reliable, self-consistent characterization of the atomic number density of Rb in the 100~mm cell, providing a robust basis for subsequent precision spectroscopy and temperature-dependent studies.\\

\begin{figure}[!h]
    \centering
    \includegraphics[width=1\linewidth]{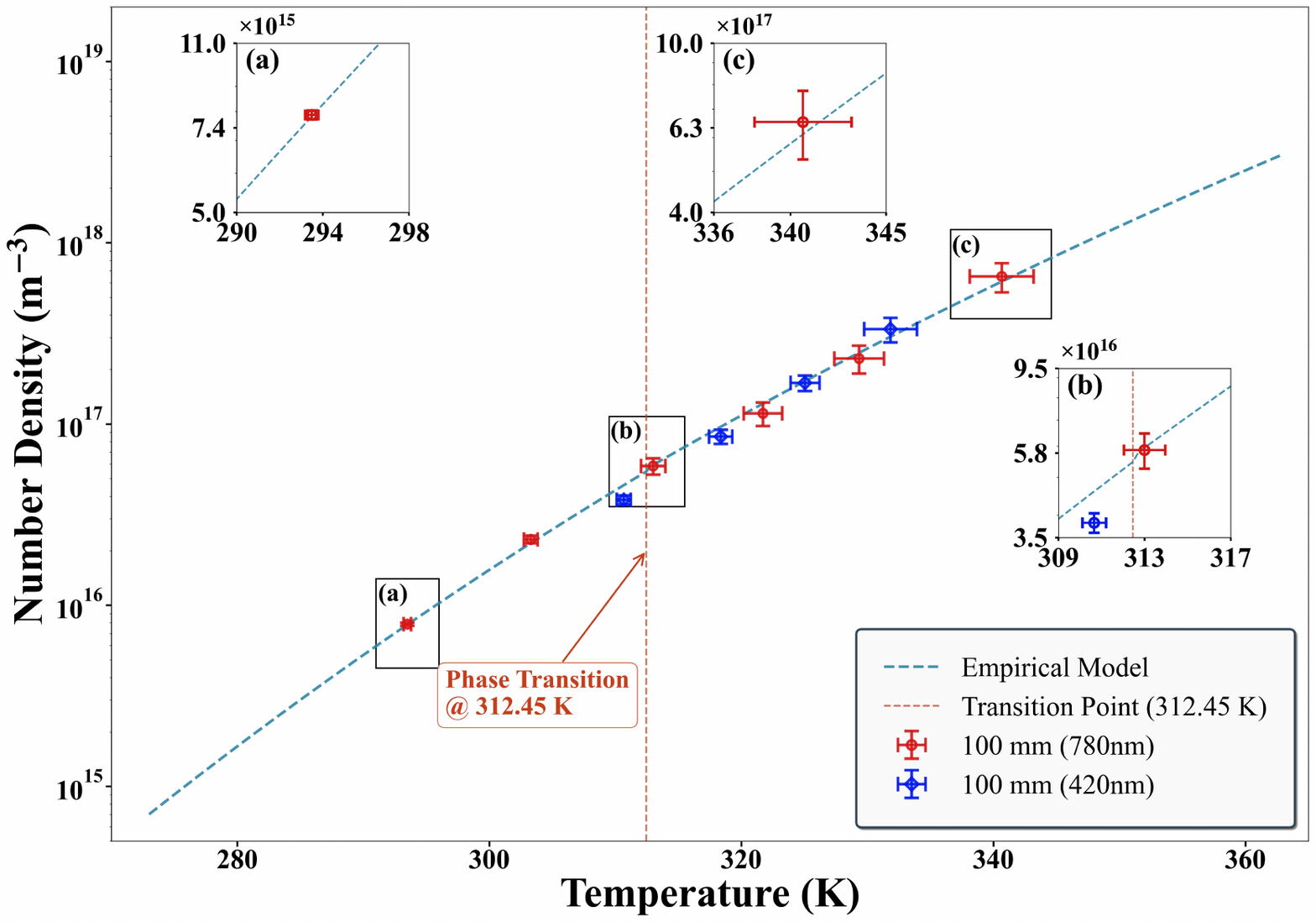}
    \caption{Atomic number density of Rb as a function of temperature for a 100 mm vapor cell. Experimental number densities extracted from absorption measurements at 780 nm (red circles) and 420 nm (blue diamonds) are compared with the empirical vapor pressure model (dashed line). The vertical dashed line marks the phase transition point at 312.45 K. Insets (a), (b), and (c) show enlarged views of the data around 293 K, 313 K, and 341 K, respectively.}
    \label{fig:number_density_100mm}
\end{figure}

\autoref{fig:number_density_100mm} shows the extracted number densities as a function of temperature obtained independently from absorption measurements at 780~nm and 420~nm, together with the empirical rubidium vapor pressure model. The two experimental datasets show strong mutual agreement and closely follow the model over the full temperature range. The vertical dashed line at 312.45~K marks the solid--liquid phase transition of rubidium, where the temperature dependence of the atomic number density changes slope, consistent with thermodynamic expectations. The insets highlight the agreement between experiment and model at selected temperatures.\\

To further validate the generality of the model, absorption spectroscopy was also performed on isotopically enriched rubidium vapor cells: an $86.5\%$ enriched $^{85}$Rb cell and a $93.6\%$ enriched $^{87}$Rb cell, each with a length of $100$~mm. For both isotopic compositions, the experimental spectra were well reproduced by the theoretical model, yielding $R^2 > 0.99$ in all cases with RMS residual below 1\%. This consistency highlights the model's versatility and reliability across different isotopic mixtures.\\

Although the NTC temperature sensors used in our experiment setup have an intrinsic accuracy of $\pm~0.3$ K, the observed error bars in the atomic number density of the 100 mm long vapor cell at higher temperatures are larger than at lower temperatures. At higher temperatures, we observe that rubidium atoms in the vapor cell tend to deposit on the inner walls through which the laser beam passes. This introduces abrupt non-linearity in the absorption signal. To mitigate this issue, we introduced a slight temperature gradient along the cell, so that the flat glass window through which the laser beam passes remains slightly hotter than the center cylindrical region. Although this configuration helps us suppress the condensation on the flat window, it marginally increases the uncertainty in the temperature measurement. 
\section{Absorption spectra at high power } \label{appnedix: high_power}

\begin{figure}[!h]
    \centering
    \includegraphics[width=1\linewidth]{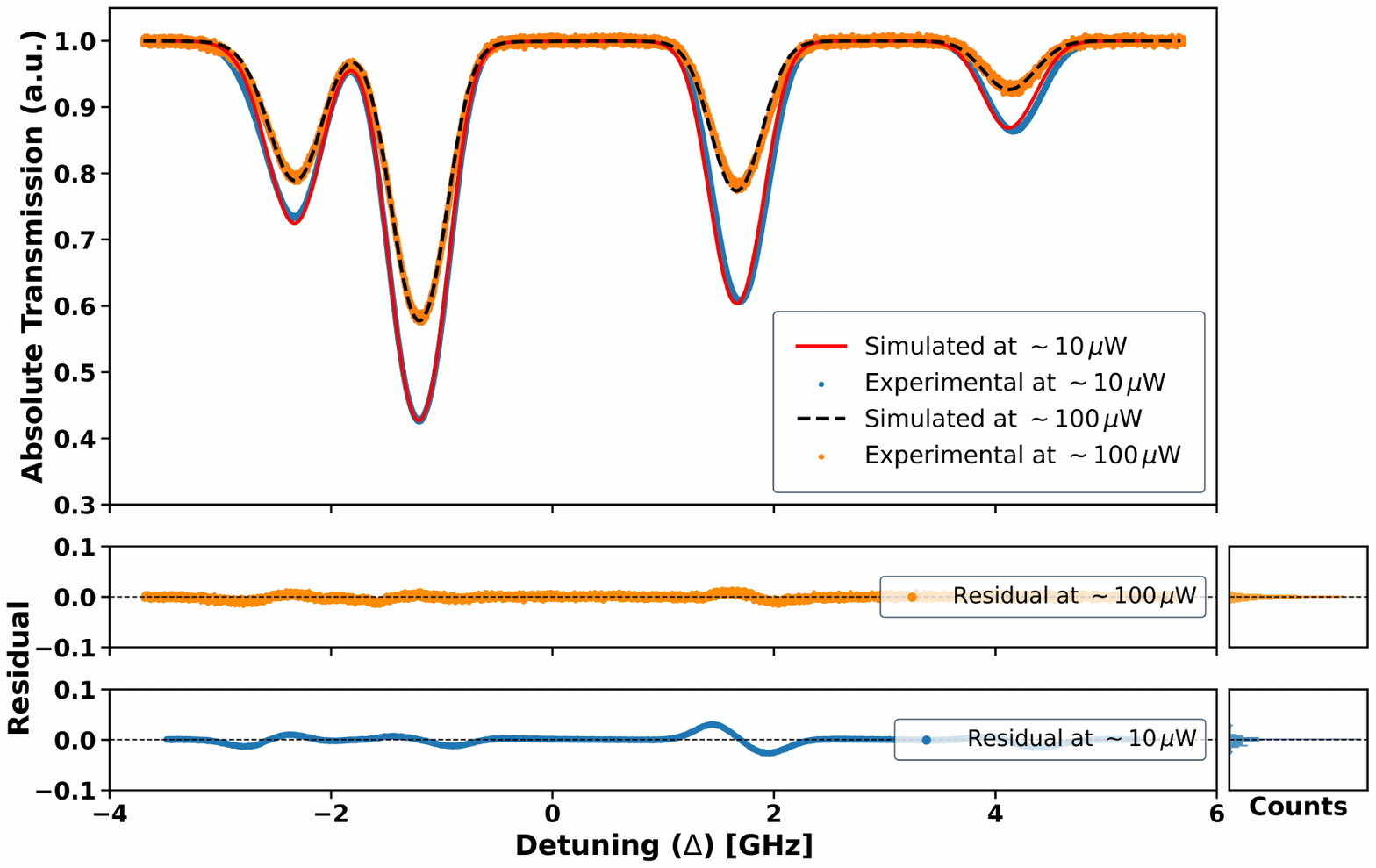}
    \caption{Comparison of experimental and theoretical transmitted laser intensity through a 100 mm rubidium vapor cell at 293.15 K for incident powers of 10~$\mu$W and 100~$\mu$W. The 10~$\mu$W data (top) represent the weak-probe regime, while the 100~$\mu$W data (bottom) show reduced absorption depth. The lower panel shows the corresponding absolute with residual histogram.}
    \label{fig:powerdependence}
\end{figure}

This section provides a detailed analysis of the high-power absorption measurements summarized in \autoref{strong-probe}. The transmission spectra were measured at probe powers of 10~$\mu$W and 100~$\mu$W and compared with simulations based on the susceptibility formalism (\autoref{fig:powerdependence}). These power levels represent distinctly different interaction regimes: the 10~$\mu$W operates in the weak-probe regime where linear absorption dominates, while the 100~$\mu$W probe power enters the nonlinear saturation regime where power broadening and reduced absorption become significant.\\

At low probe powers, both the Beer-Lambert law and the iterative susceptibility approach yield virtually identical results, as expected in the linear regime, where intensity-dependent effects are negligible. However, at higher probe powers approaching or exceeding the saturation intensity, the Beer-Lambert law becomes inadequate, since the absorption depends on the local beam intensity, which decreases along the propagation axis.\\

We used an iterative approach that accounts for intensity-dependent susceptibility along the propagation path to accurately model absorption at higher intensities. Rather than treating the vapor cell as a uniform medium, we divided the 100 mm cell into $n$ equal slices along the laser propagation axis, where each slice can be considered to have constant incident laser intensity along its length. The number of slices $n$ can be adjusted based on computational resources and desired accuracy, with more slices providing higher precision at the cost of increased computation time.\\

The intensity variation through each slice can be expressed as \cite{haupl2025modelling},
\begin{equation}
\frac{dI}{dz} = 2k* \text{Im}\left\{\sqrt{1 + \chi(I(z), \Delta, T)}\right\} I(z),
\end{equation}
where $\chi(I(z),\Delta, T)$ is the electric susceptibility, which mainly depends on the temperature (T) of the vapor cell, the detuning ($\Delta$) of the laser, and the local beam intensity ($I(z)$). This intensity dependence captures the saturation effects that become prominent at higher powers, allowing accurate modeling of the nonlinear absorption process beyond the weak-probe limit.\\

The model accurately reproduces the spectral shape and depth at both power levels. At 10~$\mu$W, the spectra exhibit the characteristic narrow linewidths expected in the weak-probe regime, whereas at 100~$\mu$W, reduced peak absorption depths are observed due to saturation effects. For our 100 mm cell length, we observed that the intensity loss through the cell remains relatively small, even at 100 $\mu$W power levels, resulting in minimal differences between the Beer-Lambert and iterative approaches. However, the iterative method becomes increasingly crucial for longer path lengths as cumulative intensity-dependent effects become more pronounced throughout the extended propagation distance. This iterative approach to absorption has also been reported in the literature, achieving results in agreement with 5\% with experiment~\cite{haupl2025modelling}.\\

The strong agreement across different probe powers, from the linear weak-probe regime to the onset of saturation, confirms the reliability and versatility of our model. This validation supports its use for extracting accurate atomic number density of alkali vapors and transition parameters beyond the weak-probe regime.\\

\section{Influence of hyperfine optical pumping}
\label{appendix:optical_pumping}

The theoretical model developed in this work incorporates hyperfine optical pumping through the branching-ratio-dependent spontaneous decay channels within the Lindblad master-equation formalism. Population redistribution between the two hyperfine ground states is taken into account. Since the extracted atomic number density is obtained by fitting the measured Doppler-broadened absorption spectrum, it is important to quantify the influence of hyperfine optical pumping on the density extraction.\\

To evaluate its significance, the experimental absorption spectrum recorded at $333.29$~K was analysed using two different theoretical models: (i) the complete model including hyperfine optical pumping, and (ii) an otherwise identical model in which the branching-ratio-induced population transfer between the two hyperfine ground states was neglected. All remaining experimental and simulation parameters were kept unchanged.\\

\begin{figure}[h!]
    \centering
    \includegraphics[width=1\linewidth]{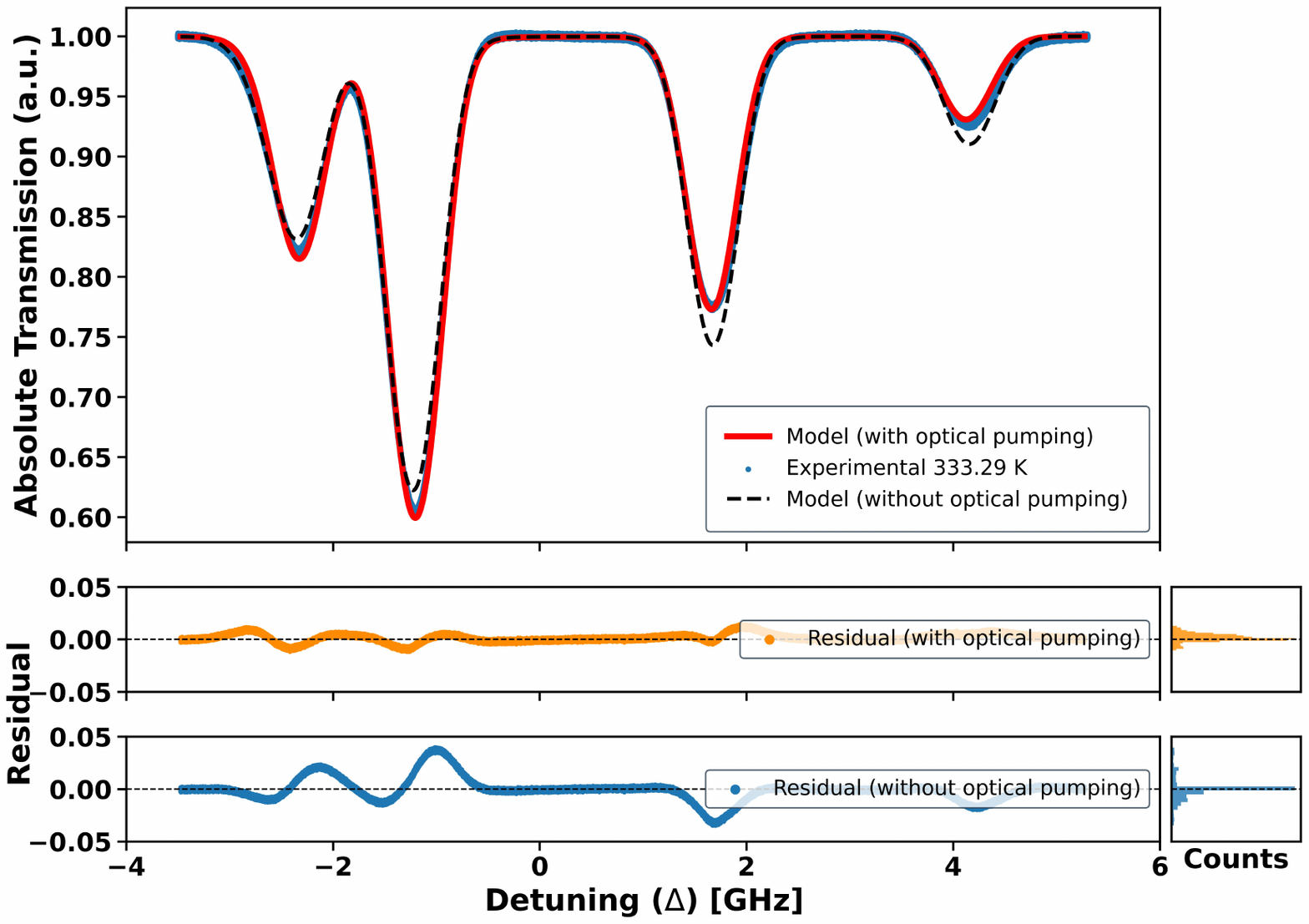}
    \caption{Comparison of the experimental absorption spectrum at $333.29$~K for the 780~nm transition measured with a 10~$\mu$W probe beam (blue dots) and the corresponding simulated spectra including hyperfine optical pumping (red solid line) and neglecting hyperfine optical pumping (dashed black line). The lower panels show the corresponding residuals (experiment$-$model). Inclusion of hyperfine optical pumping substantially improves the agreement with the experimental spectrum, reducing the RMS residual from $0.84 \%$ to $0.44 \%$ and improving the goodness of fit from $R^2=0.986$ to $R^2=0.998$. The extracted atomic number density agrees with the empirical Alcock value within $1.20\%$ when hyperfine optical pumping is included, whereas neglecting optical pumping underestimates the number density by $20.68\%$.}
    \label{fig:without_optical_pumping}
\end{figure}

\autoref{fig:without_optical_pumping} compares the experimental spectrum with the simulated spectra obtained using both models. When hyperfine optical pumping is included, the simulated spectrum accurately reproduces both the absorption profile and the relative amplitudes of the hyperfine transitions.\\

\end{document}